\documentclass[10pt]{article}

\usepackage{bm}
\usepackage{bbm}
\usepackage[final]{graphics}
\usepackage{amsmath}
\usepackage{amsfonts,amsbsy}
\usepackage{amssymb}
\usepackage{geometry}
\usepackage{titlesec}
\usepackage{subfigure}
\usepackage{amsmath}
\usepackage{float}
\usepackage{amssymb}
\usepackage{cancel}
\usepackage{xcolor}
\usepackage{slashed}
\definecolor{linkcolor}{rgb}{0.4,0.1,0.1}
\definecolor{bibcolor}{rgb}{0.4,0.1,0.1}
\usepackage[sort&compress,numbers]{natbib}
\usepackage[colorlinks,linkcolor=linkcolor,
citecolor=bibcolor,pdfpagelabels,hyperindex=true]{hyperref}
\def\empile#1\over#2{\mathrel{\mathop{\kern 0pt#1}\limits_{#2}}}

\newcommand{\slv}{\raise.15ex\hbox{$/$}\kern-.53em\hbox{$v$}}
\newcommand{\sln}{\raise.15ex\hbox{$/$}\kern-.53em\hbox{$n$}}
\newcommand{\slF}{\raise.15ex\hbox{$/$}\kern-.53em\hbox{$F$}}
\newcommand{\slL}{\raise.15ex\hbox{$/$}\kern-.53em\hbox{$L$}}
\newcommand{\slP}{\raise.15ex\hbox{$/$}\kern-.53em\hbox{$P$}}
\newcommand{\slp}{\raise.15ex\hbox{$/$}\kern-.53em\hbox{$p$}}
\newcommand{\slq}{\raise.15ex\hbox{$/$}\kern-.53em\hbox{$q$}}
\newcommand{\slR}{\raise.15ex\hbox{$/$}\kern-.53em\hbox{$R$}}
\newcommand{\slQ}{\raise.15ex\hbox{$/$}\kern-.53em\hbox{$Q$}}
\newcommand{\slK}{\raise.15ex\hbox{$/$}\kern-.53em\hbox{$K$}}
\newcommand{\slk}{\raise.15ex\hbox{$/$}\kern-.53em\hbox{$k$}}
\newcommand{\slD}{\raise.15ex\hbox{$/$}\kern-.73em\hbox{$D$}}
\newcommand{\slC}{\raise.15ex\hbox{$/$}\kern-.53em\hbox{$C$}}
\newcommand{\slA}{\raise.15ex\hbox{$/$}\kern-.53em\hbox{$A$}}
\newcommand{\slSigma}{\raise.15ex\hbox{$/$}\kern-.53em\hbox{$\Sigma$}}
\newcommand{\slpartial}{\raise.15ex\hbox{$/$}\kern-.53em\hbox{$\partial$}}
\newcommand{\slcalP}{\raise.15ex\hbox{$/$}\kern-.63em\hbox{$\cal P$}}
\newcommand{\slcalD}{\raise.15ex\hbox{$/$}\kern-.63em\hbox{$\cal D$}}

\def\p{{\boldsymbol p}}

\def\q{{\boldsymbol q}}

\def\k{{\boldsymbol k}}

\def\v{{\boldsymbol v}}

\def\u{{\boldsymbol u}}

\catcode`\@=11

\newcount\@tempcntc
\def\@citex[#1]#2{\if@filesw\immediate\write\@auxout{\string\citation{#2}}\fi
  \@tempcnta\z@\@tempcntb\m@ne\def\@citea{}\@cite{%
        \@for\@citeb:=#2\do%
    {\@ifundefined{b@\@citeb}%
        {\@citeo\@tempcntb\m@ne\@citea%
                \def\@citea{,\penalty\@m\ }{\bf ?}\@warning%
                {Citation `\@citeb' on page \thepage \space undefined}}%
        {\setbox\z@\hbox{\global\@tempcntc0\csname b@\@citeb\endcsname\relax}
     \ifnum\@tempcntc=\z@ \@citeo\@tempcntb\m@ne%
       \@citea\def\@citea{,\penalty\@m}%
       \hbox{\csname b@\@citeb\endcsname}%
     \else%
      \advance\@tempcntb\@ne%
      \ifnum\@tempcntb=\@tempcntc%
      \else\advance\@tempcntb\m@ne\@citeo%
      \@tempcnta\@tempcntc\@tempcntb\@tempcntc\fi\fi}}\@citeo}{#1}}%

\def\@citeo{\ifnum\@tempcnta>\@tempcntb\else\@citea
  \def\@citea{,\penalty\@m}%
  \ifnum\@tempcnta=\@tempcntb\the\@tempcnta\else
   {\advance\@tempcnta\@ne\ifnum\@tempcnta=\@tempcntb \else
\def\@citea{--}\fi
    \advance\@tempcnta\m@ne\the\@tempcnta\@citea\the\@tempcntb}\fi\fi}

\catcode`\@=12

\begin{document}

\title{\bf Next-to-Leading Order correction\\ 
to inclusive particle spectra in the\\ 
Color Glass Condensate framework}
\author{Fran\c cois Gelis, Julien Laidet}

\maketitle

 \begin{center}
   Institut de Physique Th\'eorique (URA 2306 du CNRS)\\
   CEA/DSM/Saclay, 91191 Gif-sur-Yvette Cedex, France
 \end{center}

\begin{abstract} 
  In [arXiv:0804.2630], we have analyzed the leading logarithms of
  energy that appear in the inclusive spectrum of gluons produced in
  heavy ion collisions, calculated in the Color Glass Condensate
  framework. The main result of this paper was that these logarithms
  are intrinsic properties of the colliding projectiles, and that they
  can be resummed by letting the distributions of color sources in the
  nuclei evolve according to the JIMWLK equation.

  An essential step in the proof of this factorization result is the
  calculation of the gluon spectrum at Next-to-Leading order, and in
  particular a functional relationship that expresses the NLO
  correction as the action of a certain operator on the LO spectrum.

  In this paper, we show that this type of relation between spectra at
  LO and NLO is not specific to the production of gluons, but that it
  is in fact generic for inclusive spectra in heavy ion collisions. To
  illustrate this, we compute up to NLO the inclusive spectrum of some
  hypothetical scalar fields, either color neutral or colored, that
  couple to gluons.
\end{abstract}

\section{Introduction}
Heavy ion collisions at high energy involve in a crucial way the
partons with a very small longitudinal momentum fraction $x$ of the
the nucleon to which they belong. For instance, in heavy ion
collisions at the LHC, where the center of mass energy is $2.76~$TeV
per nucleon pair and where the typical produced particle has a
transverse momentum of only a few GeVs, one probes partons with a
momentum fraction $x\sim 10^{-3}$ or lower. In this region, the
valence quarks are completely negligible, and most of the partons are
in fact gluons (see \cite{Aarona2} for instance). The distribution of
sea quarks--produced by the splittings $g\to q\overline{q}$--is
suppressed by a power of $\alpha_s$ compared to the gluon
distribution.

As long as the gluon occupation number remains of order one or
smaller, the evolution with $x$ of the gluon distributions is governed
by the (linear) BFKL equation \cite{BalitL1,KuraeLF1}. The evolution
of these distributions with $x$ is driven by soft gluon radiation~:
when one computes inclusive observables beyond leading order, they
contain contributions that have large logarithms of $1/x$. It turns
out that these logarithms have a universal structure, and can be
resummed simply by making the gluon distributions $x$-dependent (with
an $x$-dependence dictated by the BFKL equation). This is a particular
example of {\sl factorization}, known as $k_t$-factorization in this
case since it involves transverse momentum dependent gluon
distributions\footnote{This form of factorization has been argued to
  remain valid for some observables (namely, the single inclusive
  gluon spectrum) even when one of the projectiles has a large gluon
  occupation number, e.g. in proton-nucleus
  collisions~\cite{KovchT1,DumitM1,BlaizGV1,ChiriXY1,Avsar2,Avsar1}. But
  the transverse momentum dependent gluon distribution that describes
  the nucleus in this case is not the expectation value of the number
  operator. Instead, it has been shown that this distribution is the
  Fourier transform of a two-point correlator of Wilson lines, the
  so-called dipole amplitude.}  \cite{GunioB1,ColliE1,CatanCH1}.  It
should be emphasized that causality plays a crucial role in
factorization: it is because the two projectiles cannot be in causal
contact before they collide that their partonic content is an
intrinsic property that does not depend on what they collide with, nor
on what observable is going to be measured in the final state.

In the regime of very small values of $x$, probed in high energy heavy
ion collisions, one may reach values of the gluon occupation number
that are much larger than one. This is in fact a consequence of the
BFKL equation, that predicts a growth of the form $x^{-\lambda}$,
where $\lambda$ is some positive exponent. When the occupation number
is large, non-linear interactions among the gluons, such as
recombinations, become important~\cite{GriboLR1}. These processes tend
to stop the growth of the occupation number when it reaches a value of
the order $f\sim 1/\alpha_s$, a phenomenon known as {\sl gluon
  saturation}.  The domain where saturation effects are important is
delineated by a characteristic momentum scale, the saturation scale
$Q_s$, that depends both on $x$ and on the nucleus size $A$ (it
increases for smaller $x$ and for larger nuclei).

Computing observables in the saturation regime is a challenge for
several reasons. Firstly, since gluon recombinations are important,
processes may be initiated by more than a single gluon from each
projectile. The consequence is that in order to predict anything, one
needs informations about the distribution of multigluon states in the
wavefunctions of the two incoming projectiles. Given these multigluon
distributions, the second difficulty is to compute observables from
them. Indeed, when the gluon occupation number is of order
$\alpha_s^{-1}$, computing an observable at a given order in
$\alpha_s$ requires to resum an infinite set of Feynman
graphs. Finally, as was already the case in the dilute regime, these
observables receive large logarithmic corrections at higher
orders. But now, these logarithms may be more complicated because
their coefficients may be sensitive to the multigluon states
mentioned above. It would be highly desirable to show that these
logarithms are still universal (a property that we expect to be true
from causality), and that they can be factorized into evolved
multigluon distributions.

The Color Glass Condensate (CGC) is an effective theory, based on QCD,
in which these computations can be carried out (see
\cite{JalilKMW1,JalilKLW1,JalilKLW2,JalilKLW3,JalilKLW4,IancuLM1,IancuLM2}
for an account of the early developments and
\cite{IancuLM3,IancuV1,GelisIJV1} for reviews). The main
simplification compared to full fledged QCD stems from the realization
that the fast partons in a nucleon or nucleus have a large Lorentz
boost factor that slows down their dynamics, so that they can be
considered as static for the duration of a collision at high
energy~\cite{McLerV1,McLerV2,McLerV3}. This allows to describe them as
time independent color currents that fly along the trajectories of the
two projectiles (in the light-cone directions $z=\pm t$), whose
transverse distribution reflects the position in the transverse plane
of these fast partons at the time of the collision. All we know about
these currents is the probability distribution $W[\rho]$ for the
density $\rho_a$ of the color charges that produce them.  In contrast,
no simplification is made to the dynamics of the slower partons, and
therefore they are described as standard gauge fields, eikonally
coupled to the color current of the fast partons. The two types of
degrees of freedom (color sources and gauge fields) are separated by a
cutoff $\Lambda$ on the longitudinal momentum\footnote{In the case of
  a collision, there is one such cutoff for each nucleus: a cutoff
  $\Lambda^+$ on the momentum $k^+$ for the right moving nucleus and a
  cutoff $\Lambda^-$ for the momentum $k^-$ in the left moving
  nucleus.}.  In order to compute the expectation value of an
observable in this framework, one first computes it for a fixed
$\rho$, and then averages the result over all possible $\rho$'s
according to the probability distribution $W[\rho]$.

In the CGC framework, the power counting appropriate to the
description of the saturated regime should assume that $\rho\sim
g^{-1}$, in order to select all the relevant graphs when the gluon
occupation number is of order $1/\alpha_s$. Under this assumption, one
sees easily that observables can be expanded in powers of $g^2$, each
successive order corresponding to an additional loop. The main
difference with the $g^2$ expansion in the dilute regime is that there
are now an infinite set of graphs contributing at each order -- the
leading order (LO) corresponds to all the tree diagrams, the
next-to-leading order (NLO) to all the 1-loop graphs, etc. Moreover,
the tree diagrams that contribute to inclusive observables can be
organized in such a way that they contain only retarded propagators
\cite{GelisV2}.  Consequently, the gluon inclusive spectrum at LO can be
expressed in terms of retarded classical solutions of the Yang-Mills
equations~\cite{GelisV2}. Likewise, the quark spectrum is given at LO
by solutions of the Dirac equation in the presence of the previous
classical fields \cite{GelisKL1,GelisKL2}.

Because the CGC is an effective theory with two types of degrees of
freedom, loop integrals should be cutoff at $\Lambda^\pm$ to avoid
double counting contributions that are already included via the color
currents that describe the fast partons. However, some of these loop
integrals contain logarithms of the cutoff -- making observables
depend on an unphysical parameter that has been introduced by hand.
In the case of reactions involving a single nucleus, such as Deep
Inelastic Scattering, it is well known that these logarithms can be
absorbed into a redefinition of the probability distribution
$W[\rho]$. This redefinition amounts to letting $W$ become cutoff
dependent, with a cutoff dependence controlled by the JIMWLK
equation. When $W[\rho]$ obeys the JIMWLK equation, the cutoff
dependence in $W$ precisely cancels the cutoff dependence from the
loop integrals, leading to observables that are cutoff independent.
Moreover, it is this scale dependence of the distributions $W$ that
gives observables their rapidity dependence (for instance, the gluon
spectrum computed in the CGC framework at LO is totally independent of
rapidity, and becomes rapidity dependent after these logarithms have
been resummed).

Proving the factorization of the logarithms of the cutoffs in the case
of nucleus-nucleus collisions is more complicated. The main difficulty
is that--already at LO--observables cannot be calculated
analytically\footnote{Of course, such computations are routinely
  doable numerically by
  now~\cite{KrasnV1,KrasnV2,KrasnV3,KrasnNV1,KrasnNV2,KrasnNV4,Lappi1}.}
because this involves solving the classical Yang-Mills equations in
the presence of the two strong currents that describe the fast partons
of the two nuclei. At NLO, one needs to compute 1-loop graphs in
this classical background field, which again cannot be done
analytically. This is where causality becomes very handy~: the above
mentioned technical difficulties arise only in the forward light-cone,
i.e. after the collision has happened, while the physics that controls
the logarithms we want to compute happens before the collision, in the
wavefunctions of the two projectiles. In \cite{GelisLV3}, we have
developed a method for separating these two stages of the time
evolution in the case of the inclusive gluon spectrum, and we have
shown that this separation indeed allows one to extract analytically
the leading logarithms and to prove that they can be factorized in
JIMWLK evolved $W$ distributions (one for each nucleus). At the
moment, this factorization has been shown for inclusive
observables involving only gluons.  For instance, the computation of
the inclusive quark spectrum done in \cite{GelisKL1,GelisKL2} assumes
that such a factorization also works for quark production, although
this has not been demonstrated so far.

The cornerstone of this method is a functional relationship between
the spectrum at LO and at NLO, that makes this separation explicit
(see eq.~(\ref{eq:g-NLO}) for this formula in the case of the single
gluon spectrum). It was later found that the same separation can be
made for more complicated observables, such as the inclusive
multigluon spectra \cite{GelisLV4,GelisLV5}, or the energy-momentum
tensor, thereby allowing one to prove the factorization of their
logarithms of $\Lambda^\pm$. An important question toward extending
factorization to other observables is the range of validity of these
relationships between their tree-level and 1-loop contributions. In this
paper, we show that such formulae also exist in theories that have not
only gluons, but also other fields that couple to gluons. For
simplicity, we consider only scalar fields, but the extension of our
second example to quarks would be straightforward (apart from the
complications related to dealing with spinors rather than scalars).

In the section \ref{sec:glue}, we provide a brief reminder of existing
results concerning the gluon spectrum at LO and NLO, and the
factorization of its logarithms. More importantly, we provide a
diagrammatic method for manipulating the objects and operators that
arise in the formulae we wish to generalize. All these pictorial
representations are in a one-to-one correspondence with some
equations, but provide a much simpler and intuitive way of
manipulating them. In the section \ref{sec:scal1}, we extend this
result to some hypothetical color neutral scalar particle that couples
to a pair of gluons. Because such a field does not see directly the
background field, the results of pure Yang-Mills theory also apply
here without any change. The main part of the paper is the section
\ref{sec:scal2}, where we consider colored scalar fields that live in
the adjoint representation of the gauge group. These fields have
nontrivial interactions with the background gauge field. The main
result of this section is the eq.~(\ref{eq:theta-NLO}), a natural
generalization of the formula we had obtained for the gluon spectrum,
that relates the LO and NLO spectra. The section \ref{sec:summary} is
devoted to a summary and concluding remarks.

\section{Reminder on gluon production}
\label{sec:glue}
\subsection{Gluon spectrum at LO and NLO}
The single inclusive gluon spectrum can be expressed as follows in
terms of the color field operator $A^\mu$~:
\begin{equation}
\frac{dN_g}{d^3\p}
=
\frac{1}{(2\pi)^3 2p}\sum_{\lambda,a}
\epsilon_\mu^{(\lambda)}(\p)\epsilon_\nu^{(\lambda)*}(\p)
\int d^4x d^4y\; e^{ip\cdot(x-y)}\;
\square_x \square_y\;
\left<{A}^{\mu a}_+(x){A}^{\nu a}_-(y)\right>\; ,
\label{eq:g-all-orders}
\end{equation}
where $p\equiv |\p|$ is the on-shell energy of the produced gluon and
$\epsilon_\mu^{(\lambda)}(\p)$ is the polarization vector. In the
inclusive spectrum, one should sum over all the possible polarizations
$\lambda$ and colors $a$ of the produced gluon. The right hand side of
this formula involves the two-point $+-$ Green's function of the
Schwinger-Keldysh formalism\footnote{As opposed to a time-ordered
  Green's function in scattering amplitudes. This peculiarity can be
  seen for instance from the fact that the gluon spectrum is the
  expectation value of the number operator $a^\dagger a$.}. This is
reminded by the subscripts $+$ and $-$ carried by the two field
operators.

In the Color Glass Condensate framework, gluons are coupled to an
external classical color source $J^\mu$ that describes the fast
partons contained in the two projectiles. This means that there is a
non-zero disconnected part in the above two-point function, that must be
taken into account. Moreover, in the regime where the two projectiles
are saturated (as is the case in heavy ion collisions), one should
assume in the power counting that this source is parametrically of
order $g^{-1}$. Under these conditions, the two-point correlator can be
expanded as follows
\begin{equation}
\left<{A}^{\mu a}_+(x){A}^{\nu a}_-(y)\right>
=
\underbrace{{\cal A}^{\mu a}(x){\cal A}^{\nu a}(y)}_{\rm LO}
+
\underbrace{{\cal A}^{\mu a}(x)\alpha^{\nu a}(y)
+\alpha^{\mu a}(x){\cal A}^{\nu a}(y)
+{\cal G}_{+-}^{\mu a \nu a}(x,y)
}_{\rm NLO}
+\cdots
\end{equation}
The leading order contribution comes entirely from the disconnected
part of the two-point function, where ${\cal A}^\mu$ is the solution of
the classical Yang-Mills equations, with null boundary conditions in
the remote past,
\begin{equation}
\left[{\cal D}_\mu,{\cal F}^{\mu\nu}\right]= J^\nu\; ,\quad 
{\cal F}^{\mu\nu}\equiv\partial^\mu{\cal A}^\nu-\partial^\nu{\cal A}^\mu
+ig[{\cal A}^\mu,{\cal A}^\nu]
\; ,\quad\lim_{x^0\to-\infty}{\cal A}^\mu =0\; .
\label{eq:class-A}
\end{equation}
From this classical color field, the LO gluon spectrum is given by
\begin{equation}
\left.\frac{dN_g}{d^3\p}\right|_{_{\rm LO}}
=
\frac{1}{(2\pi)^3 2p}\sum_{\lambda,a}
\epsilon_\mu^{(\lambda)}(\p)\epsilon_\nu^{(\lambda)*}(\p)
\int d^4x d^4y\; e^{ip\cdot(x-y)}\;
\square_x \square_y\;
{\cal A}^\mu_a(x){\cal A}^\nu_a(y)\; .
\end{equation}
  The
classical field ${\cal A}^\mu$ that enters in this formula is not
known analytically in the case of a collision between two saturated
nuclei, but it is fairly straightforward to obtain it
numerically~\cite{KrasnV1,KrasnV2,KrasnV3,KrasnNV1,KrasnNV2,KrasnNV4,Lappi1}.
Note also that, at LO, the inclusive gluon spectrum is of order
\begin{equation}
\left.\frac{dN_g}{d^3\p}\right|_{_{\rm LO}}
\sim {\cal O}(g^{-2})\; ,
\end{equation}
when the two projectiles are saturated.

At next to leading order, the calculation of the gluon spectrum
involves small perturbations to the above classical field. As was
shown in \cite{GelisLV3}, these corrections can be combined into a
compact functional relationship that expresses the NLO in terms of the
LO:
\begin{eqnarray}
\left.\frac{dN_g}{d^3\p}\right|_{_{\rm NLO}}
=
\Bigg[\int\limits_\Sigma \!\!d^3\vec\u\,
\big[\alpha\cdot{\mathbbm T}_\u\big]
+
\frac{1}{2}\sum_{\lambda,a}\!\!\int\frac{d^3\k}{(2\pi)^3 2k}\!
\int\limits_\Sigma \!\!d^3\vec\u d^3\vec\v\,
\big[a_{-\k\lambda a}\cdot{\mathbbm T}_\u\big]
\big[a_{+\k\lambda a}\cdot{\mathbbm T}_\v\big]
\Bigg]\,
\left.\frac{dN_g}{d^3\p}\right|_{_{\rm LO}} .
\label{eq:g-NLO}
\end{eqnarray}
In this formula, the inclusive spectrum at leading order should be
viewed as a functional of the classical color field ${\cal A}^\mu$ on
some Cauchy surface\footnote{This surface $\Sigma$ could be any
  locally space-like surface, such that the field ${\cal A}^\mu$ above
  $\Sigma$ is completely determined by the knowledge of its value and
  that of its first time derivative on $\Sigma$. $d^3\vec\u$ is the
  3-volume measure on this surface.} $\Sigma$. The fields $\alpha$ and
$a_{\pm\k\lambda a}$ are small perturbations to the classical color
field on $\Sigma$, that are defined in detail in \cite{GelisLV3}. The
operator ${\mathbbm T}_\u$ is the generator of the shifts\footnote{The
  retarded classical color field ${\cal A}^\mu$ implicitly depends on
  its initial condition ${\cal A}_{_\Sigma}$ on the initial Cauchy
  surface $\Sigma$. Any functional $F[{\cal A}]$ can thus be viewed as
  a functional $G[{\cal A}_{_\Sigma}]$ of the initial condition. The
  value of this functional for the new classical field resulting from
  a shift of the initial condition can be obtained by exponentiating
  the operator ${\mathbbm T}_\u$,
\begin{equation*}
  G[{\cal A}_{_\Sigma}+\alpha]
=
\exp\Big[\int\limits_\Sigma d^3\vec\u\;[\alpha\cdot{\mathbbm T}_\u]\Big]\;
G[{\cal A}_{_\Sigma}]\; .
\end{equation*} (this relationship can be seen as a definition of ${\mathbbm T}_\u$.)} of the classical field ${\cal A}$ at the point
$\u\in\Sigma$.

\subsection{Leading logarithms and factorization}
The operator in the square brackets in the right hand side of
eq.~(\ref{eq:g-NLO}) contains momentum integrals -- the explicit
integration over $d^3\k$ in the second term, and a loop integral
hidden in the definition of $\alpha$. These integrals are both
divergent when the longitudinal components of the momentum, $k^+$ or
$k^-$, go to infinity. In the CGC framework, these integrations should
have upper limits $\Lambda^\pm$, where $\Lambda^\pm$ is the cutoff
that separates the fields from the classical sources, in order to
prevent double countings. The $k^\pm$ integrals are thus finite, but
they give logarithms of the unphysical cutoffs $\Lambda^\pm$. In
\cite{GelisLV3}, the following result was established
\begin{eqnarray}
&&
\int\limits_\Sigma \!\!d^3\vec\u\,
\big[\alpha\cdot{\mathbbm T}_\u\big]
+
\frac{1}{2}\sum_{\lambda,a}\!\!\int\frac{d^3\k}{(2\pi)^3 2k}\!
\int\limits_\Sigma \!\!d^3\vec\u d^3\vec\v\,
\big[a_{-\k\lambda a}\cdot{\mathbbm T}_\u\big]
\big[a_{+\k\lambda a}\cdot{\mathbbm T}_\v\big]
=\nonumber\\
&&
\qquad\qquad\qquad\qquad\qquad
=
\log(\Lambda^+)\,{\cal H}_1
+
\log(\Lambda^-)\,{\cal H}_2
+\mbox{terms w/o logs}\; ,
\label{eq:NLO-logs}
\end{eqnarray}
where the operators ${\cal H}_1$ and ${\cal H}_2$ are the JIMWLK
Hamiltonians of the two nuclei. The most important aspect of this
formula is that the logarithms of the two cutoffs are multiplied by
objects that depend only on the sources of the corresponding nucleus.
This property is crucial for the universality --and thus the
factorization-- of these logarithms: it ensures that the logarithms of
$\Lambda^+$ are an intrinsic property of the projectile moving in the
$+z$ direction, that does not depend in any way on the projectile
moving in the opposite direction.

Thanks to eq.~(\ref{eq:NLO-logs}), one can resum all the leading
powers of the logarithms (i.e. all the terms of the form $(\alpha_s
\log\Lambda^\pm)^n$), and absorb them into the scale dependence of
the distributions $W_1[\rho_1]$ and $W_2[\rho_2]$ that describe the
source content of the two projectiles:
\begin{equation}
\left.\frac{dN_g}{d^3\p}\right|_{\mbox{\rm Leading Log}}
=
\int\big[D\rho_1 D\rho_2\big]\; W_1[\rho_1]W_2[\rho_2]\;
\left.\frac{dN_g}{d^3\p}\right|_{_{\rm LO}} ,
\label{eq:g-LLog}
\end{equation}
where the two distributions $W$ obey the JIMWLK equation
\begin{equation}
\Lambda^\pm\frac{\partial W_{1,2}}{\partial\Lambda^\pm}
=
-{\cal H}_{1,2}\,W_{1,2}\; .
\end{equation}

\subsection{Diagrammatic manipulations with ${\mathbbm T}_\u$}
\label{sec:diagT}
In the above NLO expression (\ref{eq:g-NLO}), a very compact result is
obtained thanks to the introduction of the operator ${\mathbbm T}_\u$.
This operator will play a crucial role in the rest of this paper, and
for this reason it is very useful to gain some intuition on how it
acts on various objects. All the formulae involving ${\mathbbm T}_\u$
can be proven by making use of the Green's formulae that relate
classical fields --and perturbations thereof-- to their initial value on
the surface $\Sigma$. However, these proofs are often cumbersome (see
\cite{GelisLV3,GelisLV4} for some examples). In this subsection, we
would like to present a more intuitive, diagrammatic, way of
manipulating these operators.

First of all, let us recall the most basic identity (derived in
\cite{GelisLV3}), that relates a linearized perturbation $a^\mu$
propagating over the classical background field ${\cal A}^\mu$ to the
classical field itself
\begin{eqnarray}
a^\mu(x)
=
\int\limits_\Sigma d^3\vec\u\;[a\cdot{\mathbbm T}_\u]\;{\cal A}^\mu(x)\; .
\label{eq:T-basic}
\end{eqnarray}
This formula means that if we know the perturbation on the surface
$\Sigma$ and how the classical field ${\cal A}(x)$ depends on its
initial condition on $\Sigma$, then we can obtain the value of the
perturbation at the point $x$ by acting on ${\cal A}(x)$ with the
operator ${\mathbbm T}_\u$. Roughly speaking, ${\mathbbm T}_\u$ acts
as a first order derivative with respect to the initial value of
${\cal A}$ (and that of its first time derivative) at the point
$\u\in\Sigma$. Diagrammatically, the identity of
eq.~(\ref{eq:T-basic}), before the integration over $d^3\vec\u$ is
performed, can be represented as follows~: \setbox1\hbox to
5mm{\hfil\resizebox*{!}{12mm}{\includegraphics{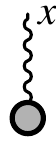}}} \setbox2\hbox to
10mm{\hfil\resizebox*{!}{12mm}{\includegraphics{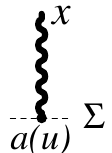}}}
\begin{equation}
[a\cdot{\mathbbm T}_\u]\;\;\raise -6mm\box1\;
=\;\raise -5.5mm\box2\;\; .
\label{eq:T-A}
\end{equation}
In the left hand side, the thin wavy line terminated by a gray blob
represents the retarded classical color field at the point $x$. In the
right hand side, one has a perturbation that starts on $\Sigma$ as
$a(\u)$, and then propagates to the point $x$ over the classical
background (the boldface propagator thus indicates that it is dressed
by this background field). In this diagrammatic representation of
eq.~(\ref{eq:T-basic}), one should keep in mind the following rule~:
{\sl when the combination $[a\cdot{\mathbbm T}_\u]$ acts on some
  object, whatever is attached to the operator ${\mathbbm T}_\u$ (here
  $a(\u)$) replaces the part of that object that hangs below $\Sigma$
  at the point $\u\in\Sigma$. }

Let us now consider the action of an operator that contains several
powers of ${\mathbbm T}$, the simplest of which is $[a\cdot{\mathbbm
  T}_\u][b\cdot{\mathbbm T}_\v]$, that involves two ${\mathbbm T}$'s at
different points $\u,\v\in\Sigma$. When this combination acts on the
retarded classical field ${\cal A}^\mu$, one gets an object that
involves the dressed 3-point vertex with $a(\u)$ and $b(\v)$
attached to two of its endpoints:
\setbox1\hbox to
5mm{\hfil\resizebox*{!}{12mm}{\includegraphics{A}}} \setbox2\hbox to
10mm{\hfil\resizebox*{!}{18mm}{\includegraphics{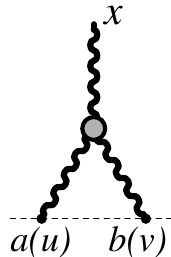}}}
\begin{equation}
[a\cdot{\mathbbm T}_\u][b\cdot{\mathbbm T}_\v]\;\;\raise -6mm\box1\;
=\;\raise -9mm\box2\quad\;\; .
\label{eq:TT-A}
\end{equation}
In the right hand side of this relationship, the blob indicates that
the 3-gluon vertex is dressed by the background field,
\setbox1\hbox to
50mm{\hfil\resizebox*{!}{13mm}{\includegraphics{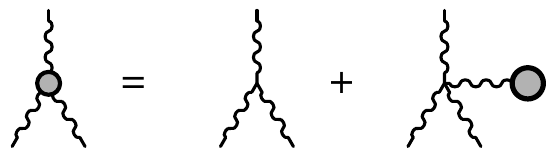}}}
\begin{equation}
\raise -6mm\box1\;\;,
\end{equation}
i.e. it is the sum of the bare 3-gluon vertex and a 4-gluon vertex
where one of the four legs is attached to the classical background field.

In order to understand diagrammatically the equation (\ref{eq:g-NLO}),
we need also the following representations,\setbox1\hbox to
13mm{\hfil\resizebox*{!}{16mm}{\includegraphics{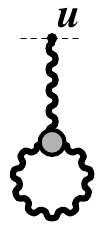}}}\setbox2\hbox to
19mm{\hfil\resizebox*{!}{9mm}{\includegraphics{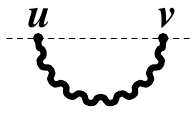}}}
\begin{eqnarray}
\alpha(\u)&=&\raise -8mm\box1
\nonumber\\
\sum_{\lambda,a}\int\frac{d^3\k}{(2\pi)^3 2k}\;
a_{-\k\lambda a}(\u)
a_{+\k\lambda a}(\v)&=&\raise -4mm\box2
\; .
\label{eq:alpha}
\end{eqnarray}
The first one is just the definition of $\alpha(\u)$ as the 1-loop
correction to the field expectation value at the point $\u\in\Sigma$.
The second equation is a representation of the $G_{+-}$ propagator
(dressed by the background field) as a sum over a complete basis of
fluctuations that propagate over this background (see \cite{GelisLV3},
and also eq.~(\ref{eq:G+--LO}) later in this paper). It is now easy to
stitch eqs.~(\ref{eq:T-A}), (\ref{eq:TT-A}) and (\ref{eq:alpha}), in
order to get\setbox1\hbox to
5mm{\hfil\resizebox*{!}{12mm}{\includegraphics{A}}}\setbox2\hbox to
11mm{\hfil\resizebox*{!}{18mm}{\includegraphics{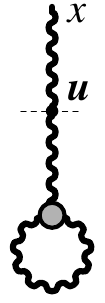}}}\setbox3\hbox
to
17mm{\hfil\resizebox*{!}{18mm}{\includegraphics{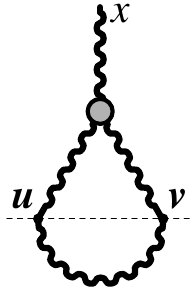}}}\setbox4\hbox
to 14mm{\hfil\resizebox*{!}{18mm}{\includegraphics{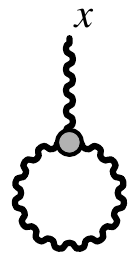}}}
\begin{eqnarray}
&&
\Bigg[\int\limits_\Sigma \!\!d^3\vec\u\,
\big[\alpha\cdot{\mathbbm T}_\u\big]
+
\frac{1}{2}\sum_{\lambda,a}\!\!\int\frac{d^3\k}{(2\pi)^3 2k}\!
\int\limits_\Sigma \!\!d^3\vec\u \,d^3\vec\v\,
\big[a_{-\k\lambda a}\cdot{\mathbbm T}_\u\big]
\big[a_{+\k\lambda a}\cdot{\mathbbm T}_\v\big]
\Bigg]\;\raise -5.5mm\box1=
\nonumber\\
&&\qquad=\int\limits_\Sigma d^3\vec\u\raise -10mm\box2
+
\frac{1}{2}\int\limits_\Sigma d^3\vec\u \,d^3\vec\v\raise -10mm\box3
=\raise -9mm\box4\; .
\end{eqnarray}
The two terms in the left hand side of the second line correspond to
the two possible localizations of the 3-gluon vertex: below or
above\footnote{A crucial aspect in all these manipulations is the fact
  that the inclusive quantities can always be expressed in terms of
  retarded propagators, that describe the causal propagation of some
  object over a background field. Therefore, the ``stitching''
  procedure described above simply corresponds to concatenating the
  evolution from $-\infty$ to the surface $\Sigma$ (encoded in
  the prefactors of the ${\mathbbm T}$'s) and the evolution from
  $\Sigma$ to the time $x^0$ of interest (encoded in the
  object upon which the ${\mathbbm T}$'s act).} the surface
$\Sigma$. When combined, these two terms  reconstruct the
complete 1-loop correction to the expectation value of the field
at the point $x$ (the prefactor $1/2$ in the second term gives the
correct symmetry factor for the tadpole).

Since ${\mathbbm T}_\u$ acts as a derivative, its action on a product
of classical fields is distributed according to Leibnitz's
rule. Since the gluon spectrum at LO is the Fourier transform of the
product ${\cal A}(x){\cal A}(y)$, it is instructive to see how
${\mathbbm T}_\u$ acts on this product.  For instance, on
gets\setbox1\hbox to
16mm{\hfil\resizebox*{!}{12mm}{\includegraphics{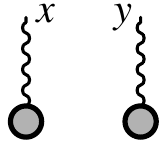}}} \setbox2\hbox to
16mm{\hfil\resizebox*{!}{12mm}{\includegraphics{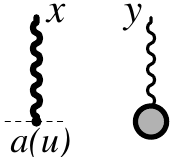}}}\setbox3\hbox
to 16mm{\hfil\resizebox*{!}{12mm}{\includegraphics{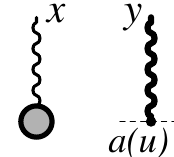}}}
\begin{equation}
[a\cdot{\mathbbm T}_\u]
\;\;\Bigg[\raise -5.5mm\box1\;\Bigg]
=\;\raise -5.5mm\box2\;\;+\raise -5.5mm\box3\;\;,
\end{equation}
\setbox4\hbox to 16mm{\hfil\resizebox*{!}{12mm}{\includegraphics{A2}}}%
\setbox5\hbox to 16mm{\hfil\resizebox*{!}{12mm}{\includegraphics{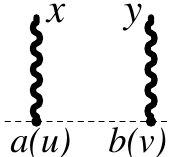}}}%
\setbox6\hbox to 16mm{\hfil\resizebox*{!}{12mm}{\includegraphics{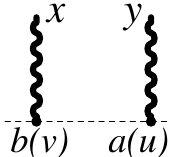}}}%
\setbox2\hbox to 16mm{\hfil\resizebox*{!}{17mm}{\includegraphics{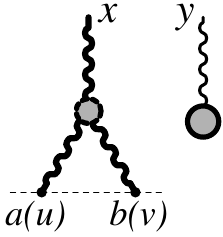}}}%
\setbox3\hbox to 16mm{\hfil\resizebox*{!}{17mm}{\includegraphics{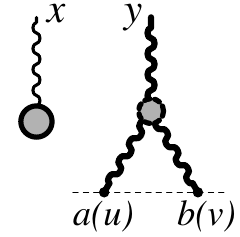}}}%
and
\begin{equation}
[a\cdot{\mathbbm T}_\u]
[b\cdot{\mathbbm T}_\v]
\;\;\Bigg[\raise -5.5mm\box4\;\Bigg]
=\;\raise -5.5mm\box5\;\;\;+\raise -5.5mm\box6
\;+\raise -10mm\box2\;\;\;+\raise -10mm\box3\;.
\end{equation}
Combining these two equations with eqs.~(\ref{eq:alpha}), one sees
graphically that eq.~(\ref{eq:g-NLO}) generates exactly the
terms that are needed to obtain the gluon spectrum at NLO,%
\setbox1\hbox to 16mm{\hfil\resizebox*{!}{12mm}{\includegraphics{A2}}}%
\setbox2\hbox to 16mm{\hfil\resizebox*{!}{15mm}{\includegraphics{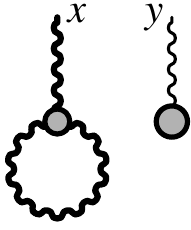}}}%
\setbox3\hbox to 16mm{\hfil\resizebox*{!}{15mm}{\includegraphics{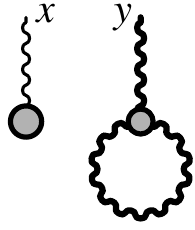}}}%
\setbox4\hbox to 16mm{\hfil\resizebox*{!}{15mm}{\includegraphics{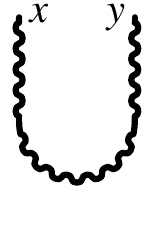}}}%
\begin{eqnarray}
&&
\Bigg[\int\limits_\Sigma \!\!d^3\vec\u\,
\big[\alpha\cdot{\mathbbm T}_\u\big]
+
\frac{1}{2}\sum_{\lambda,a}\!\!\int\frac{d^3\k}{(2\pi)^3 2k}\!
\int\limits_\Sigma \!\!d^3\vec\u \,d^3\vec\v\,
\big[a_{-\k\lambda a}\cdot{\mathbbm T}_\u\big]
\big[a_{+\k\lambda a}\cdot{\mathbbm T}_\v\big]
\Bigg]\;\Bigg[\raise -5.5mm\box1\;\Bigg]=
\nonumber\\
&&\qquad\qquad\qquad\qquad
=
\raise -9mm\box2
\;\;+
\raise -9mm\box3
\;\;+
\raise -9mm\box4
\;\; .
\end{eqnarray}
This diagrammatic approach is very powerful, since the above
manipulations are arguably much simpler than the derivation of the NLO
gluon spectrum in \cite{GelisLV3}. Alternatively, if one does not go
all the way to relying only on these diagrams, one can use them as a
guidance to guess some identities, that can then be proven by the more
analytical  (and tedious) methods of \cite{GelisLV3}.

\section{Color neutral scalar particles}
\label{sec:scal1}
\subsection{Model}
The simplest extension to the pure Yang-Mills case is to consider the
production of some hypothetical scalar particle, singlet under the
gauge group of QCD, that couples to the square of the gluon field
strength. Such a particle can be described by a field $\phi$ whose
interactions are given in the following Lagrangian density,
\begin{equation}
{\cal L}\equiv \underbrace{\frac{1}{2}(\partial_\mu\phi)(\partial^\mu\phi) 
-\frac{1}{2}m^2 \phi^2}_{\mbox{free scalar field}}
-\underbrace{\frac{1}{4}g_{\phi}\;\phi\,F_{\mu\nu}^a F^{\mu\nu,a}}_{\mbox{coupling to gluons}}\; .
\label{eq:L-phi}
\end{equation}
For instance, this particle could be a Higgs boson, and this
Lagrangian would be an effective description of its coupling to a pair
of gluons. At the fundamental level, this coupling occurs mostly via a
top quark loop, and one would obtain this Lagrangian by integrating
out the quarks, leading to an effective coupling constant $g_\phi$.
Note that this model was considered recently in \cite{Liou1}, in the
limit where the scalar particle mass is much larger than the
saturation momentum. This limit is simpler because the leading term
(when we expand both in $g^2$ and $Q_s^2/m^2$) in the production
amplitude has one $F^{\mu\nu}$ attached to each of the colliding
nuclei, with no entanglement between the sources of the two nuclei,
leading to a much simpler form of factorization. In this section, we
consider a generic mass $m^2$ for this scalar particle, possibly as
small as the saturation scale, and therefore we do not expand
observables in $Q_s^2/m^2$.

Here, we disregard the self-interactions of these scalar particles,
and consider only their coupling to gluons. Moreover, in the power
counting, we assume that the effective coupling constant $g_\phi$ is
much smaller than the strong coupling constant $g$. Therefore, we only
consider contributions that have the lowest possible order in
$g_\phi$, and the perturbative expansion is in powers of $g$. This
means that the NLO corrections affect the gluons, and only indirectly
the scalar field $\phi$ via its coupling to $F^2$.

In this model, one can see the gluon field strength squared,
multiplied by the coupling $g_\phi$, as a source for the scalar
$\phi$,
\begin{equation}
J\equiv -\frac{1}{4}g_{\phi}\,F_{\mu\nu}^a F^{\mu\nu,a}\; .
\end{equation}
The Lagrangian (\ref{eq:L-phi}) is thus that of a free scalar field
coupled to the external source $J$.

\subsection{Inclusive spectrum at Leading Order}
In close analogy with the gluon spectrum (\ref{eq:g-all-orders}), the
single inclusive spectrum for $\phi$ is given by the following
formula,
\begin{equation}
\frac{dN_\phi}{d^3\p}
=
\frac{1}{(2\pi)^3 2E_\p}\int d^4x d^4y\; e^{ip\cdot(x-y)}\;
(\square_x+m^2)(\square_y+m^2)
\left<\phi_+(x)\phi_-(y)\right>\; ,
\label{eq:phi-allO}
\end{equation}
in terms of the two-point $+-$ function for the field $\phi$. This
formula is true to all orders, but at leading order in $g^2$ the
two-point correlator that appears under the integral is equal to the
product $\varphi(x)\varphi(y)$, where $\varphi(x)$ is the retarded
solution of the classical equation of motion for $\phi$,
\begin{equation}
(\square_x+m^2)\,\varphi(x)=
-\frac{1}{4}g_{\phi}\,{\cal F}_{\mu\nu}^a {\cal F}^{\mu\nu,a}
\quad,\quad\lim_{x^0\to-\infty}\varphi(x)=0\; .
\end{equation}
In this equation, it is sufficient to keep only the leading order
contribution to the gluon field strength, i.e. its classical value
given by eq.~(\ref{eq:class-A}).  In the regime where the colliding
projectiles have a saturated gluon content, the gluon field strength
is of order $1/g$. Thus the source $J$ is of order $g_\phi/g^2$, and
the leading contribution to the $\phi$ spectrum is of order
$g_\phi^2/g^4$. In contrast, the first connected contribution to
$\big<\phi_+(x)\phi_-(y)\big>$ starts only at the order
$g_\phi^2/g^2$.

\begin{figure}[htbp]
\begin{center}
\resizebox*{!}{1.7cm}{\includegraphics{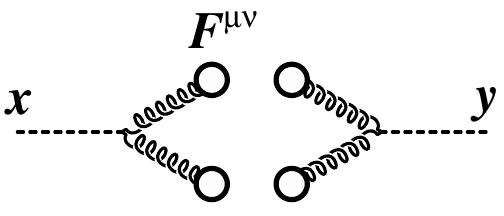}}
\end{center}
\caption{\label{fig:phi-LO}Diagrammatic representation of the
  contributions to the inclusive $\phi$ spectrum at Leading Order. The
  dotted line is the free propagator of the scalar particle
  (eliminated in the spectrum by the action of the operators
  $\square+m^2$). Each spiraling line terminated by a circular blob
  represents one instance of the gluon field strength at tree level
  (i.e. it contains terms of order one and two in the gauge field ${\cal
    A}^\mu$).}
\end{figure}

Thus, at LO, the $\phi$ spectrum is a sum of tree diagrams (see the
figure \ref{fig:phi-LO}).  In terms of the classical field strength,
the inclusive spectrum for $\phi$ at LO reads simply
\begin{equation}
\left.\frac{dN_\phi}{d^3\p}\right|_{_{\rm LO}}
=
\frac{1}{(2\pi)^3 2E_\p}\frac{g^2_\phi}{16}\int d^4x d^4y\; e^{ip\cdot(x-y)}\;
{\cal F}_{\mu\nu}^a(x){\cal F}^{\mu\nu,a}(x)\;
{\cal F}_{\rho\sigma}^b(y){\cal F}^{\rho\sigma,b}(y)\; .
\end{equation}
This formula simply expresses the well known fact that, when
non-self-interacting particles are produced by an external source, the
inclusive spectrum is the modulus squared of the Fourier transform of
the source.

\subsection{Next-to-Leading Order correction}
\begin{figure}[htbp]
\begin{center}
\resizebox*{!}{1.6cm}{\includegraphics{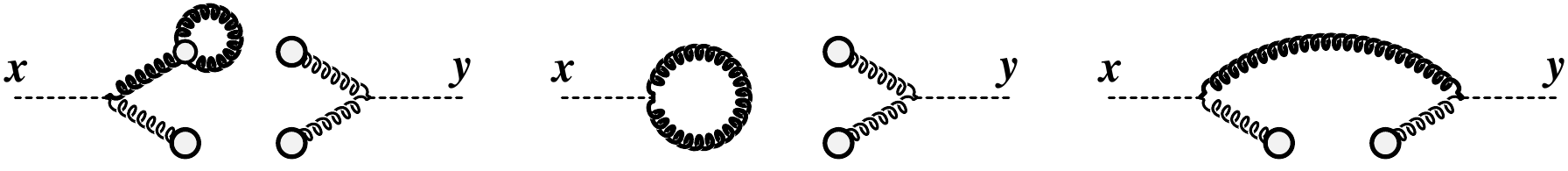}}
\end{center}
\caption{\label{fig:phi-NLO}Diagrammatic representation of the
  contributions to the inclusive $\phi$ spectrum at Next-to-Leading
  Order. Each blob represents an infinite sum of tree level
  graphs. The diagram on the left contains the 1-loop correction to
  $F^{\mu\nu}(x)$. The diagram in the middle involves the connected
  part of the 1-loop correction to $F^{\mu\nu}(x)F_{\mu\nu}(x)$ (there
  are two other topologies, not represented here, where the 1-loop
  correction is on $\Phi_-(y)$). The right diagram contains the
  lowest order connected contribution to
  $F^{\mu\nu}(x)F^{\rho\sigma}(y)$. Thick lines indicate that
  the gluon propagators are dressed by the classical background field
  ${\cal A}^\mu$.}
\end{figure}

Let us now consider the production of the $\phi$-particles at NLO,
i.e. at the next order in the strong coupling constant $g^2$. At the
next order, one has two types of contributions:
\begin{itemize}
\item[{\bf i.}] 1-loop corrections to the gluon field strength, or to the
  gluon field strength squared.
\item[{\bf ii.}] the connected part of $\big<\phi_+(x)\phi_-(y)\big>$
  at tree level.
\end{itemize}
These two contributions are illustrated in the figure \ref{fig:phi-NLO}.
These corrections affect only the gluons, and they can be handled in
exactly the same way as the corrections to the inclusive gluon spectrum.

The manipulations performed in the previous section for the
expectation value of $A^\mu(x) A^\nu(y)$ can be extended to this more
complicated situation, leading the following formal
expression\footnote{One can easily see that this derivation applies to
  any quantity that can be expressed at leading order as a local
  function of the retarded classical color field. It also applies to
  bi-local expressions --i.e. involving the gauge field at two points
  $x$ and $y$-- provided one of these points carries the
  Schwinger-Keldysh index $+$ and the other one the index $-$. For
  more than two points or more general assignments of the
  Schwinger-Keldysh indices, it still works provided all the pairs of
  points have space-like separations.} for the inclusive $\phi$
spectrum at NLO,
\begin{eqnarray}
\left.\frac{dN_\phi}{d^3\p}\right|_{_{\rm NLO}}
=
\Bigg[\int\limits_\Sigma \!\!d^3\vec\u\,
\big[\alpha\cdot{\mathbbm T}_\u\big]
+
\frac{1}{2}\sum_{\lambda,a}\!\!\int\frac{d^3\k}{(2\pi)^3 2k}\!
\int\limits_\Sigma \!\!d^3\vec\u d^3\vec\v\,
\big[a_{-\k\lambda a}\cdot{\mathbbm T}_\u\big]
\big[a_{+\k\lambda a}\cdot{\mathbbm T}_\v\big]
\Bigg]\,
\left.\frac{dN_\phi}{d^3\p}\right|_{_{\rm LO}} .
\label{eq:phi-NLO}
\end{eqnarray}
In this formula, the inclusive spectrum at leading order should again
be considered as a functional of the classical color field ${\cal
  A}^\mu$ on some Cauchy surface $\Sigma$. This relationship is
formally identical to the one that relates purely gluonic observables
at LO and NLO (e.g. inclusive gluon spectra, or the gluon contribution
to the energy-momentum tensor).

In the gluon case, we have shown that it is the integration over $\k$
in the right hand side of eq.~(\ref{eq:phi-NLO}) that leads to the
logarithms of the CGC cutoffs. Since we have here exactly the same
operator, the logarithms that appear in the NLO correction to the
inclusive $\phi$ spectrum are identical to those encountered in the
gluon spectrum, and they can be resummed into the JIMWLK evolution of
the distributions of the sources that produce the color field,
\begin{equation}
\left.\frac{dN_\phi}{d^3\p}\right|_{\mbox{\rm Leading Log}}
=
\int\big[D\rho_1 D\rho_2\big]\; W_1[\rho_1]W_2[\rho_2]\;
\left.\frac{dN_\phi}{d^3\p}\right|_{_{\rm LO}} .
\end{equation}
This is another example of the universality of these logarithms.

\section{Adjoint scalars}
\label{sec:scal2}
\subsection{Model}
The simplicity of the previous example is to a large extent due to the
fact that the field $\phi$ is not charged under the gauge group of
strong interactions. Therefore, the NLO corrections are entirely due
to corrections to the gauge fields themselves, and to their two-point
correlations.

Things get more complicated when the produced particle carries a color
charge. An obvious example is of course that of quarks and
antiquarks. However, since quarks also bring the complication of being
described by anticommuting fermion fields, let us consider as an
intermediate step the case of some hypothetical colored scalar field
$\theta_a$, that lives in the adjoint representation of the gauge
group. The Lagrangian for such scalars is
\begin{equation}
{\cal L}\equiv \big(D_\mu\theta\big)_a\big(D^\mu\theta\big)^*_a\; ,
\label{eq:theta-L}
\end{equation}
where $D_\mu$ is the covariant derivative that ensures the minimal
coupling of the scalars to the gluons. We assume here that these
scalar fields do not interact directly among themselves (i.e. their
interactions are always mediated by gluons).

\subsection{Inclusive spectrum at LO}
The starting point to compute the single inclusive spectrum for the
$\theta$ particles is the analogue of eq.~(\ref{eq:phi-allO}),
\begin{equation}
\frac{dN_\theta}{d^3\p}
=
\frac{1}{(2\pi)^3 2p}\sum_c\int d^4x d^4y\; e^{ip\cdot(x-y)}\;
\square_x\square_y\;
\big<\theta_+^c(x)\theta_-^c(y)\big>
\; ,
\label{eq:theta-allO}
\end{equation}
where $c$ is the color of the produced scalar particle\footnote{In the
  following, when the context is not ambiguous, we may omit the color
  indices attached to the endpoints of the objects we are manipulating,
  in order to keep the notations lighter.} (summed over in the
definition of the inclusive spectrum).  The two-point correlator under
the integral differs from the previous two examples in the fact that
it has only connected contributions. This is due to the fact that the
Lagrangian (\ref{eq:theta-L}) is even in the field $\theta$. To stress
this property, let us rewrite eq.~(\ref{eq:theta-allO}) as follows
\begin{equation}
\frac{dN_\theta}{d^3\p}
=
\frac{1}{(2\pi)^3 2p}\sum_c\int d^4x d^4y\; e^{ip\cdot(x-y)}\;
\square_x\square_y\;D_{+-}^{cc}(x,y)\; ,
\label{eq:theta-allO1}
\end{equation}
where $D_{+-}^{bc}$ is a connected two-point function.  The Leading Order
contribution to $D_{+-}^{bc}$, that we will denote ${\cal D}_{+-}^{bc}$, is the
sum of all the tree level graphs, i.e. the two-point function dressed by
the classical color field ${\cal A}^\mu$, solution of
eq.~(\ref{eq:class-A}).
\begin{figure}[htbp]
\begin{center}
\resizebox*{!}{0.6cm}{\includegraphics{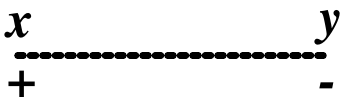}}
\end{center}
\caption{\label{fig:scalar-LO}Diagrammatic representation of the only
  contribution to the inclusive $\theta$ spectrum at Leading Order
  (disconnected graphs cannot exist because the Lagrangian is even in
  $\theta$). The thick line indicate that the propagator is dressed by
  insertions of the classical color field ${\cal A}^\mu$.}
\end{figure}
${\cal D}_{+-}^{bc}$ can be expressed in terms of a complete basis of
solutions of the scalar wave equation on top of the classical color
field ${\cal A}^\mu$,
\begin{eqnarray}
  {\cal D}_{+-}^{bc}(x,y)
  &=&
  \sum_a\int\frac{d^3\q}{(2\pi)^3 2q}\;
  \vartheta_{-\q a}^b(x)\vartheta_{+\q a}^c(y)
  \nonumber\\
  \left({\cal D}_\mu{\cal D}^\mu\right)_{cb}\,\vartheta_{\pm\q a}^b(x) &=& 0\; ,\quad \lim_{x^0\to -\infty}\vartheta_{\pm\q a}^b(x) = \delta_{ab} e^{\pm i q \cdot x}\; .
\label{eq:D+--LO}
\end{eqnarray}
In the notation $\vartheta_{\pm\q a}^b$(x), one should not confuse the two
color indices $a$ and $b$. The lower index $a$ indicates the color of
the wave in the remote past, before it propagates over the classical
color field, while the upper index $b$ indicates the color at the
point $x$.
The fields $\vartheta_{\pm\q a}$ can be represented diagrammatically as follows~:
\setbox1\hbox to 2.3cm{\hfil\resizebox*{!}{2cm}{\includegraphics{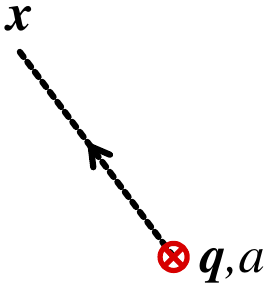}}}
\begin{equation}
\vartheta_{\pm\q a}(x) = \raise -8mm\box1\; .
\end{equation}
In this representation, the red cross represents the initial condition
at $x^0\to-\infty$, i.e. a plane wave with fixed momentum and color.
The arrow is a reminder of the fact that the functions
$\vartheta_{\pm\q a}$ obey retarded boundary conditions set at $x^0\to
-\infty$. In other words, the propagator that connects the red cross
to the point $x$ is a retarded propagator (dressed by the classical
color field ${\cal A}^\mu$).

In terms of these objects, one can represent the LO propagator ${\cal
  D}_{+-}$ as \setbox1\hbox to
2.8cm{\hfil\resizebox*{!}{2cm}{\includegraphics{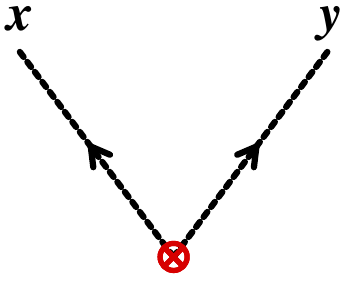}}}
\begin{equation}
{\cal D}_{+-}(x,y) = \raise -8mm\box1\; ,
\label{eq:D+--LO1}
\end{equation}
and the $\theta$ spectrum at LO is given by
\begin{equation}
\left.\frac{dN_\theta}{d^3\p}\right|_{_{\rm LO}}
=
\frac{1}{(2\pi)^3 2p}\sum_{a,c}\int\frac{d^3\q}{(2\pi)^3 2q}\int d^4x d^4y\; e^{ip\cdot(x-y)}\;
\square_x\square_y\;\vartheta_{-\q a}^c(x)\vartheta_{+\q a}^c(y)\; .
\label{eq:theta-LO}
\end{equation}
In this formula for the $\theta$ spectrum at leading order, $\q$ can
be interpreted as the momentum of the antiparticle that must
necessarily be produced along with the particle of momentum $\p$ that
is tagged in the measurement.

\subsection{NLO corrections}
\subsubsection{List of the 1-loop corrections}
\setbox1\hbox to 6.5cm{\hfil\resizebox*{!}{1.7cm}{\includegraphics{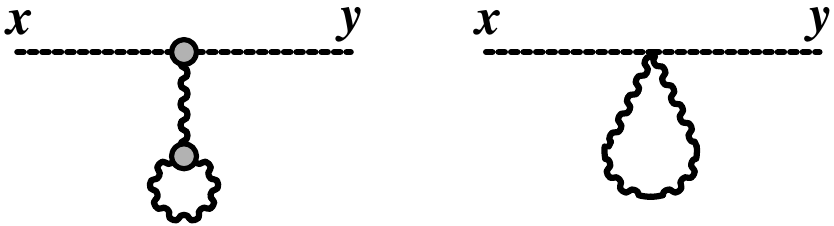}}\hfil}
\setbox2\hbox to 4.1cm{\hfil\resizebox*{!}{1.7cm}{\includegraphics{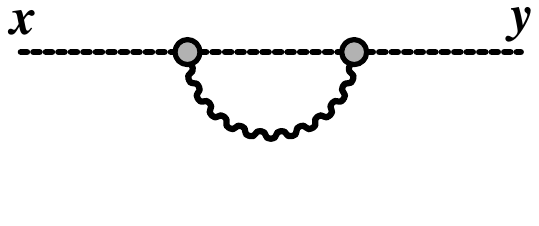}}\hfil}
\begin{figure}[htbp]
\begin{center}
$$\underbrace{\raise 1mm\box1}_{D_{+-}^{(a)}(x,y)}\qquad
\underbrace{\raise 1mm\box2}_{D_{+-}^{(b)}(x,y)}$$
\end{center}
\caption{\label{fig:scalar-NLO}Diagrammatic representation of the
  contributions to the inclusive $\theta$ spectrum at Next-to-Leading
  Order.  All the lines represent propagators dressed by insertions of
  the classical color field ${\cal A}^\mu$. The blobs indicate that
  the corresponding vertices are dressed by the background field.}
\end{figure}
The NLO corrections to the inclusive spectrum of the $\theta$
particles are represented in the figure\footnote{Note that the
  $gg\theta\theta^*$ vertex is not dressed at tree level.}
\ref{fig:scalar-NLO}. They can be divided in two distinct categories:
\begin{itemize}
\item[{\bf i.}] Tadpole insertions on the propagator of the $\theta$
  field, that appear in the first two diagrams of the figure
  \ref{fig:scalar-NLO}. The leftmost correction is the 1-loop
  correction $\alpha^\mu$ to the classical color field ${\cal A}^\mu$
  that enters in the right hand side of eq.~(\ref{eq:g-NLO}). The
  second diagram is a local self-energy correction on the propagator
  of the $\theta$ field. 
\item[{\bf ii.}] Non-local 1-loop self-energy correction to the scalar
  propagator, represented in the third diagram of the figure
  \ref{fig:scalar-NLO}. 
\end{itemize}

\subsubsection{Alternate representation}
\label{sec:alt}
Our goal for the rest of this paper is to write the NLO corrections to
the $\theta$ spectrum in a form similar to eqs.~(\ref{eq:g-NLO}) and
(\ref{eq:phi-NLO}). This will be done in several steps. In this
subsection, we first rewrite these corrections in a form that
resembles eq.~(\ref{eq:D+--LO1}), in order to highlight the terms
that can be seen as mere corrections to one of the functions
$\vartheta_{\pm\q a}$.

This is very simple in the case of the two terms on the left of the
figure \ref{fig:scalar-NLO}. Indeed, in these two terms the
self-energy insertion is a tadpole, that by definition corrects the
propagator at a single point. This implies that one can see them as a
correction to one of the two factors $\vartheta_{\pm\q a}$.  Let us
denote $\zeta_{\pm\q a}$ this ${\cal O}(g^2)$ correction to
$\vartheta_{\pm\q a}$, \setbox1\hbox to
43mm{\hfil\resizebox*{4.3cm}{!}{\includegraphics{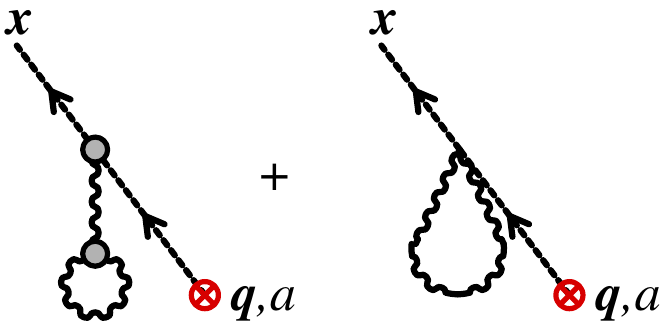}}}
\begin{equation}
\zeta_{\pm\q a}^b(x)
=
\raise -9mm\box1\; .
\label{eq:zeta1}
\end{equation}
In terms of $\zeta_{\pm\q a}$, the corresponding NLO corrections to
the propagator $D_{+-}$ reads
\begin{equation}
D_{+-}^{(a)}(x,y)
=
\sum_a\int\frac{d^3\q}{(2\pi)^3 2k}\;
  \left[\vartheta_{-\q a}(x)\zeta_{+\q a}(y)
+\zeta_{-\q a}(x)\vartheta_{+\q a}(y)\right]\; .
\end{equation}
This equation simply states that each of the two factors
$\vartheta_{\pm\q a}$ in eq.~(\ref{eq:D+--LO}) has to be corrected in
turn. Diagrammatically, this can be represented as
\setbox1\hbox to 53mm{\hfil\resizebox*{5.3cm}{!}{\includegraphics{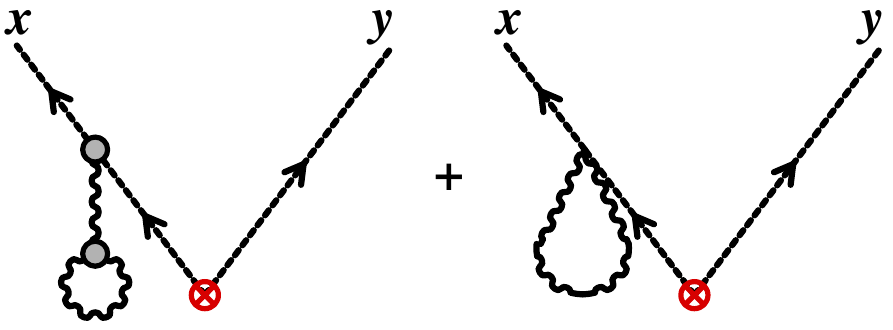}}}
\begin{equation}
D_{+-}^{(a)}(x,y)
=
\raise -9mm\box1\; +[x\leftrightarrow y]\;,
\label{eq:Da}
\end{equation}
where the unwritten terms, hidden under the abbreviation
$[x\leftrightarrow y]$, are the two terms where the tadpole is
attached to the leg that ends at the point $y$.

Let us now consider the third diagram on the right of the figure
\ref{fig:scalar-NLO}. This term is the $+-$ scalar propagator,
corrected by a non-local 1-loop self-energy insertion. Because the
self-energy is a genuine two-point function, this correction cannot be
seen as a correction of either $\vartheta_{+\q a}$ or $\vartheta_{-\q
  a}$ in eq.~(\ref{eq:D+--LO}). Instead, we must go back to the
Schwinger-Keldysh formalism in order to evaluate this term.

Let us call $\Sigma$ the self-energy. Summing over all the possible
assignments of the internal $\pm$ indices in the Schwinger-Keldysh
formalism, we can first write
\begin{equation}
D_{+-}^{(b)}
=
{\cal D}_{++} \Sigma_{++} {\cal D}_{+-}
-
{\cal D}_{++}\Sigma_{+-}{\cal D}_{--}
-
{\cal D}_{+-}\Sigma_{-+}{\cal D}_{+-}
+
{\cal D}_{+-}\Sigma_{--}{\cal D}_{--}\; ,
\end{equation}
where the ${\cal D}_{\pm\pm}$ are the four components of the scalar
propagator at tree level. To keep the notations simple, we have not
written explicitly the space-time integrations involved in
concatenating the self-energy and the tree-level propagators in this
equation, but we have of course included the appropriate signs to keep
track of the fact that the vertices of type $-$ are opposite to those
of type $+$. After some simple algebra, one can rearrange the four
terms in the right hand side of the previous equation as follows
\begin{equation}
D_{+-}^{(b)}
=
{\cal D}_{_R} \Sigma_{_R} {\cal D}_{+-}
+
{\cal D}_{_R}\Sigma_{+-}{\cal D}_{_A}
+
{\cal D}_{+-}\Sigma_{_A}{\cal D}_{_A}\; ,
\label{eq:Db}
\end{equation}
where the retarded and advanced propagators and self-energies are
related to the Schwinger-Keldysh ones by
\begin{eqnarray}
{\cal D}_{_R}&\equiv&{\cal D}_{++}-{\cal D}_{+-}={\cal D}_{-+}-{\cal D}_{--}
\nonumber\\
{\cal D}_{_A}&\equiv&{\cal D}_{++}-{\cal D}_{-+}={\cal D}_{+-}-{\cal D}_{--}
\nonumber\\
\Sigma_{_R}&\equiv&\Sigma_{++}-\Sigma_{+-}=\Sigma_{-+}-\Sigma_{--}
\nonumber\\
\Sigma_{_A}&\equiv&\Sigma_{++}-\Sigma_{-+}=\Sigma_{+-}-\Sigma_{--}\; .
\end{eqnarray}
It is interesting to note the pattern\footnote{This pattern would
  remain true even when more than three propagators and self-energies
  are concatenated.} in eq.~(\ref{eq:Db})~: each term in this formula
contains exactly one object of type $+-$. Everything on the left of
this object is retarded, and everything on its right is advanced.  The
three terms of eq.~(\ref{eq:Db}) exhaust all the possible locations of
the object that carries the $+-$ indices.

For the factors ${\cal D}_{+-}$, we can use eq.~(\ref{eq:D+--LO}).  In
coordinate space, the 1-loop self-energy $\Sigma_{+-}$ is the product
of a scalar and a gluon propagator~:
\begin{equation}
\Sigma_{+-}\propto {\cal D}_{+-} {\cal G}_{+-}\; ,
\end{equation}
where the symbol $\propto$ means that we have not written the vertices
(they are a bit cumbersome to write because of the possibility of
attaching the background field ${\cal A}^\mu$ to one leg of the
$gg\theta\theta^*$ vertex to form an effective $g\theta\theta^*$
vertex -- how the complete vertices arise will become clear later).  In
this formula ${\cal G}_{+-}$ is the gluon propagator dressed by
insertions of the classical gauge field ${\cal A}^\mu$. There is for
${\cal G}_{+-}$ a formula identical to eq.~(\ref{eq:D+--LO}),
\begin{equation}
{\cal G}_{+-}(x,y)
  =
  \sum_{a,\lambda}\int\frac{d^3\k}{(2\pi)^3 2k}\;
  a_{-\k\lambda a}(x)a_{+\k\lambda a}(y)
  \; ,
\label{eq:G+--LO}
\end{equation}
where the $a_{\pm\k\lambda a}$ are now gluonic fluctuations that
propagate on top of ${\cal A}^\mu$, starting as plane waves of
momentum $\k$, polarization $\lambda$ and color $a$ when
$x^0\to-\infty$ (see \cite{GelisLV3} for more details). Mimicking
eq.~(\ref{eq:D+--LO1}), we represent this diagrammatically as
follows~:
\setbox1\hbox to 2.8cm{\hfil\resizebox*{!}{2cm}{\includegraphics{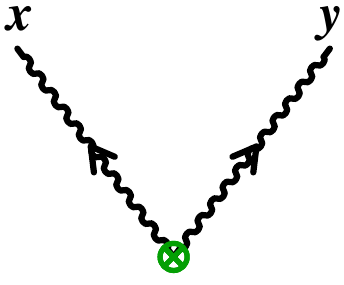}}}
\begin{equation}
{\cal G}_{+-}(x,y) = \raise -8mm\box1\; .
\label{eq:G+--LO1}
\end{equation}
The three terms of eq.~(\ref{eq:Db}) can therefore be represented in
the following way,
\setbox1\hbox to 2.3cm{\hfil\resizebox*{!}{1.8cm}{\includegraphics{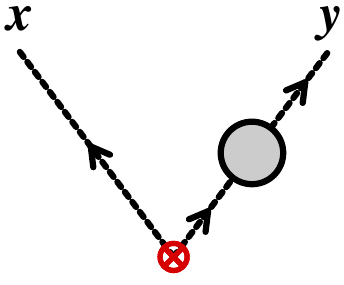}}}
\setbox2\hbox to 2.3cm{\hfil\resizebox*{!}{1.8cm}{\includegraphics{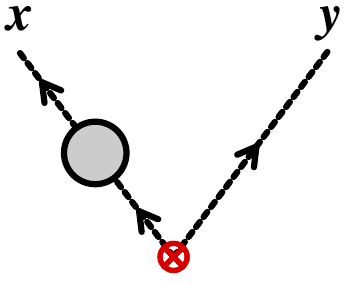}}}
\setbox3\hbox to 2.3cm{\hfil\resizebox*{!}{1.8cm}{\includegraphics{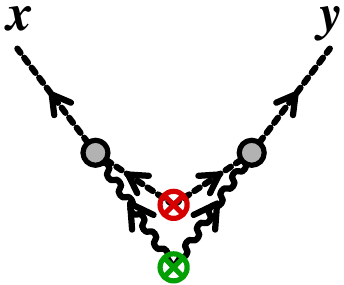}}}
\begin{eqnarray}
&&
{\cal D}_{_R}\Sigma_{_R}{\cal D}_{+-}= \raise -8mm\box2\quad ,\quad
{\cal D}_{+-}\Sigma_{_A}{\cal D}_{_A}= \raise -8mm\box1\nonumber\\
&&
{\cal D}_{_R}\Sigma_{+-}{\cal D}_{_A}= \raise -8mm\box3\; ,
\end{eqnarray}
where the big grey blobs represent the self-energies $\Sigma_{_{R,A}}$.

At this point, it is convenient to rewrite the retarded
self-energy\footnote{A similar reorganization of the terms can be
  performed on the advanced one.} $\Sigma_R$. First of all, one can
write
\begin{eqnarray}
\Sigma_{_R}
&=&
\Sigma_{++}-\Sigma_{+-}
\nonumber\\
&\propto&
{\cal G}_{++}{\cal D}_{++}-{\cal G}_{+-}{\cal D}_{+-}
\nonumber\\
&=&
{\cal G}_{_R}{\cal D}_{++}+{\cal G}_{+-}{\cal D}_{_R}
\nonumber\\
&=&
{\cal G}_{_R}{\cal D}_{-+}+{\cal G}_{+-}{\cal D}_{_R}\; .
\label{eq:SigR1}
\end{eqnarray}
In order to obtain the last equality, we must recall that
\begin{equation}
{\cal D}_{++}(x,y)
=
\theta(x^0-y^0)\,{\cal D}_{-+}(x,y)
+
\theta(y^0-x^0)\,{\cal D}_{+-}(x,y)\; .
\end{equation}
Since this propagator is multiplied by the retarded propagator ${\cal
  G}_{_R}$, the terms with $\theta(y^0-x^0)$ is forbidden, and we can
in fact replace the ${\cal D}_{++}$ by ${\cal D}_{-+}$ to get the
final formula.  
In this rearrangement, we have chosen to break the symmetry between
the ${\cal D}$ and ${\cal G}$ propagators in a certain way. It is
possible to obtain a more symmetric formula. By doing slightly
different transformations, we could alternatively obtain
\begin{eqnarray}
\Sigma_{_R}
\propto
{\cal D}_{_R}{\cal G}_{-+}+{\cal D}_{+-}{\cal G}_{_R}\; ,
\label{eq:SigR2}
\end{eqnarray}
and, by combining eqs.~(\ref{eq:SigR1}) and (\ref{eq:SigR2}),
\begin{eqnarray}
\Sigma_{_R}
\propto
\frac{1}{2}{\cal D}_{_R}({\cal G}_{-+}+{\cal G}_{+-})
+
\frac{1}{2}({\cal D}_{+-}+{\cal D}_{-+}){\cal G}_{_R}\; .
\label{eq:SigR3}
\end{eqnarray}
Diagrammatically, the expression given in
eq.~(\ref{eq:SigR3}) for the retarded self-energy can be represented
as follows,
\setbox1\hbox to 4.3cm{\hfil\resizebox*{!}{1.8cm}{\includegraphics{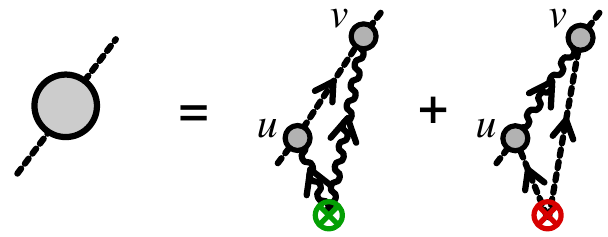}}}
\begin{equation}
\raise -9mm\box1\qquad .
\label{eq:SigR4}
\end{equation}
This representation also highlights the
different physical meanings of the two terms:
\begin{itemize}
\item In the first term, a vacuum fluctuation made of a pair of
  virtual gluons is created in the past (at the location of the green
  cross). One of these gluons is absorbed by the scalar fluctuation at
  the point $u$, and the second of these gluons is absorbed by the
  same scalar a bit later, at the point $v$.
\item In the second term, the vacuum fluctuation created at the
  location of the red cross is a $\theta\theta^*$ fluctuation. One of
  the scalars from this vacuum fluctuation annihilates the incoming
  scalar at the point $u$ to form a gluon. Later, at the point $v$,
  this gluon is absorbed by the second scalar from the vacuum
  fluctuation to give the outgoing line.
\end{itemize}

\subsection{Expression of the NLO corrections in terms of the operators
  ${\mathbbm T}$}
\subsubsection{Introduction}
Our goal is now to relate the NLO corrections (at the exception of
those that involve $\theta\theta^*$ vacuum fluctuations -- see later)
to the LO spectrum by means of a formula similar to
eqs.~(\ref{eq:g-NLO}) and (\ref{eq:phi-NLO}). In order to keep the
derivation as intuitive as possible, we will perform this graphically,
by using the rules discussed in the section \ref{sec:diagT}. Before we
proceed, let us summarize the previous discussion by listing all the
terms involved in the $\theta$ spectrum at NLO~:
\setbox1\hbox to 13cm{\hfil\resizebox*{!}{4.0cm}{\includegraphics{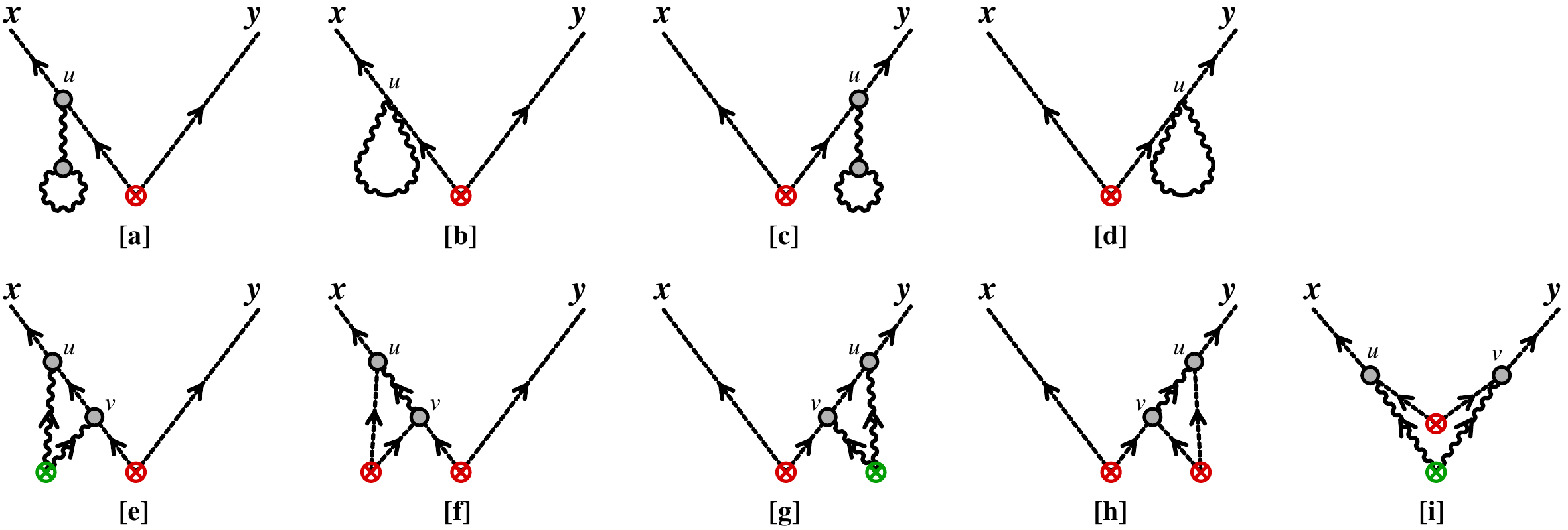}}}
\begin{equation}
\raise -10mm\box1\qquad .
\label{eq:NLO-all}
\end{equation}
In these graphs, the red and green crosses are objects that are pinned
at $x^0=-\infty$ (because they represent the initial condition in the
remote past for the scalar waves $\vartheta_{\pm\q a}$ and gluonic
waves $a_{\pm \k\lambda a}$). The other vertices in these diagrams can
be located anywhere in space-time, and the corresponding formulae
contain integrals over the positions of all these vertices.  In
particular, these vertices can be located above or below the Cauchy
surface $\Sigma$.  The only constraint on their locations is that they
are {\sl ordered} by the causal property of the retarded propagators:
the endpoint of a retarded propagator must lie in the future
light-cone of its starting point.

\subsubsection{Action of the gluonic shifts}
As a first step, let us determine the action on the LO spectrum
of the operator 
\begin{equation}
\int\limits_\Sigma \!\!d^3\vec\u\,
\big[\alpha\cdot{\mathbbm T}_\u\big]
+
\frac{1}{2}\sum_{\lambda,a}\!\!\int\frac{d^3\k}{(2\pi)^3 2k}\!
\int\limits_\Sigma \!\!d^3\vec\u d^3\vec\v\,
\big[a_{-\k\lambda a}\cdot{\mathbbm T}_\u\big]
\big[a_{+\k\lambda a}\cdot{\mathbbm T}_\v\big]\; ,
\end{equation}
that appears in eqs.~(\ref{eq:g-NLO}) and (\ref{eq:phi-NLO}). The
${\mathbbm T}$'s contained in this operator act only on the initial
condition on $\Sigma$ for the gauge field (implicitly hidden in all
the propagators, since they are all dressed by the classical
background field). Let us start with the action of $[a\cdot{\mathbbm
  T}_\u]$ on the LO spectrum. Graphically, we have \setbox1\hbox to
5cm{\hfil\resizebox*{!}{1.8cm}{\includegraphics{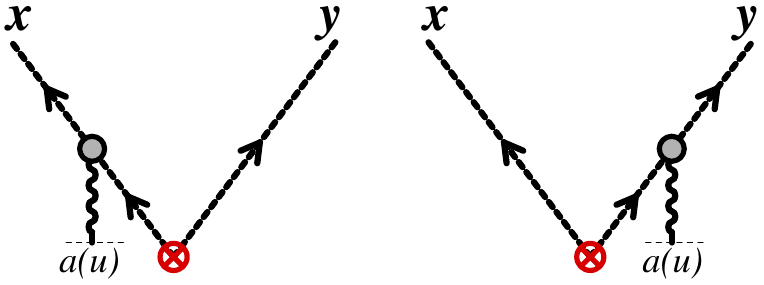}}}
\begin{equation}
[a\cdot{\mathbbm T}_\u]\;{\cal D}_{+-}(x,y)
=
\raise -9mm\box1\; .
\end{equation}
It is equally straightforward to write the action of the quadratic
operator $[a\cdot{\mathbbm T}_\u][b\cdot{\mathbbm T}_\v]$ on the
leading order expression, \setbox1\hbox to
9cm{\hfil\resizebox*{!}{6.3cm}{\includegraphics{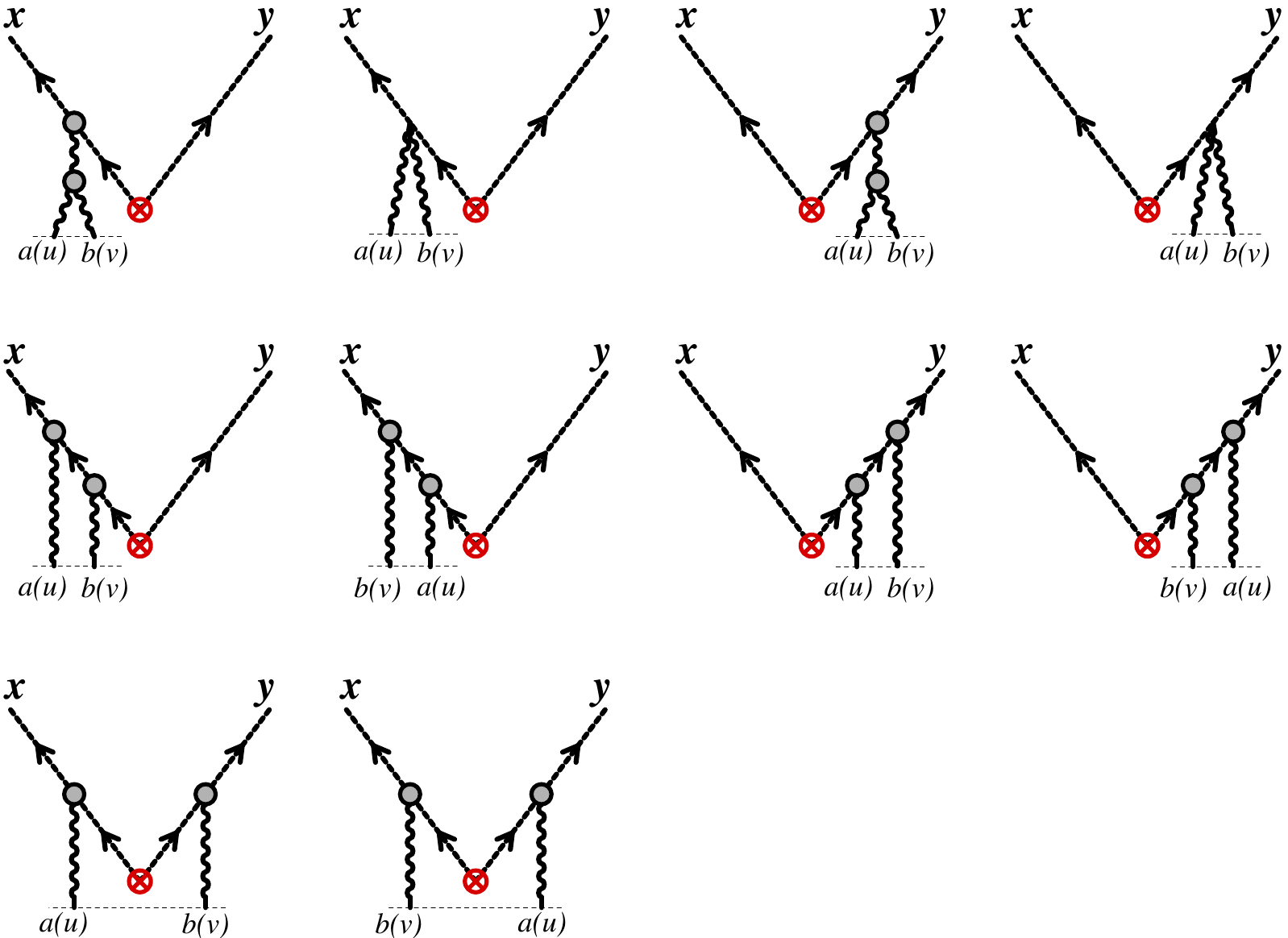}}}
\begin{equation}
[a\cdot{\mathbbm T}_\u][b\cdot{\mathbbm T}_\v]\;{\cal D}_{+-}(x,y)
=
\raise -29mm\box1\; .
\end{equation}

When we contract $\alpha(u)$ into ${\mathbbm T}_\u$ and
$\frac{1}{2}\sum_{\lambda, a}\int_\k a_{-\k\lambda a}(u)a_{+\k\lambda
  a}(v)$ into ${\mathbbm T}_\u {\mathbbm T}_\v$, and integrate over
$u,v\in\Sigma$, we generate a good deal of the NLO terms in
eq.~(\ref{eq:NLO-all}). However, by inspection of the graphs that are
generated in this operation and by comparing with the list in
eq.~(\ref{eq:NLO-all}), we readily see that two types of terms are
missing:
\begin{itemize}
\item[{\bf i.}] The graphs {\bf [f,h]}, where the NLO correction is
  due to a $\theta\theta^*$ vacuum fluctuation, are missing entirely.
  Having in mind to use this NLO calculation in order to extract the
  leading logs of energy, we are going to disregard these terms from
  now on. Indeed, as in the derivation of the BFKL equation, only the
  gluonic vacuum fluctuations contribute at leading logarithmic
  accuracy. Scalar or fermionic fluctuations do not contribute to the
  logarithms at this order, and become necessary only when one goes at
  the next to leading logarithmic accuracy.
\item[{\bf ii.}] In the graphs {\bf [a,b,c,d,e,g,i]}, only the portion
  of the integration domain where the vertices $u$ and $v$ are both
  above the surface $\Sigma$ are generated. The missing pieces of
  these graphs are important even at leading log, and in the rest of
  this section we will show how to generate them.
\end{itemize}

\subsubsection{Scalar shifts}
In the terms of eq.~(\ref{eq:NLO-all}), when one of the vertices $u$
or $v$ (or both) is located below the surface $\Sigma$, it means that
the scalar waves $\vartheta_{\pm\q a}$ have been affected by the
gluonic fluctuation before they reach $\Sigma$. This implies that the
corresponding modification to the final observable should be generated
by operators that act on the value of the scalar fields on $\Sigma$,
instead of the operator ${\mathbbm T}_\u$ that acts on the gluon
fields on $\Sigma$.

In order to see this, let us introduce the analogue of the operator
${\mathbbm T}_\u$ for the fields $\vartheta_{\pm\q a}$. Consider a
generic solution of the following equation of motion
\begin{equation}
\left({\cal D}_\mu{\cal D}^\mu\right)_{cb}\,\vartheta^b(x)=0\; ,
\end{equation}
that describes the propagator of a scalar over some gauge background
(The gauge field in the background being fixed.) The Green's formula
for the solutions of this equation reads
\begin{eqnarray}
\vartheta^b(x)
&=&
i\int\limits_\Omega d^4y\; D_{_R}^0(x,y)\;\Big[
2ig{\cal A}_\mu(y)\partial^\mu_y
+
ig(\partial_y^\mu{\cal A}_\mu(y))
+
g^2{\cal A}_\mu(y){\cal A}^\mu(y)
\Big]^{bc}\,\vartheta^c(y)
\nonumber\\
&&\qquad
+
i\int\limits_\Sigma d^3\vec{\u}\;
D_{_R}^0(x,u)\;(n\cdot\stackrel{\leftrightarrow}{\partial}_u)\,
\vartheta^b(u)\; ,
\end{eqnarray}
where $D_{_R}^0$ is the free retarded propagator for the scalar field
$\theta$ (i.e. the retarded Green's function of the D'Alembertian
operator $\square_x$). The domain $\Omega$ for the integration in the
first line is the region of space-time located above the surface
$\Sigma$. The term in the second line is a boundary term that contains
all the dependence of the solution on its initial condition on
$\Sigma$ ($n^\mu$ is the normalized vector normal to $\Sigma$ at the
point $u$).

Let us now define the following operator,
\begin{equation}
\xi\cdot{\mathbbm T}^{\vartheta}_\u
\equiv
\xi^b(u)\frac{\delta}{\delta\vartheta^b(u)}
+(n\cdot\partial\xi^b(u))\frac{\delta}{\delta(n\cdot\partial\vartheta^b(u))}
\; ,
\label{eq:T-theta}
\end{equation}
that mimics for scalar fields the operator ${\mathbbm T}_\u$ that was
introduced for gauge fields. Diagrammatically, the action of this
operator can be pictured as follows,
\setbox1\hbox to 1.7cm{\resizebox*{!}{20mm}{\hfil\includegraphics{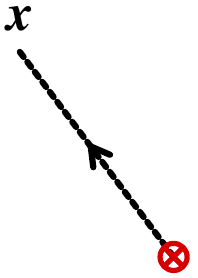}}}
\setbox2\hbox to 1.8cm{\resizebox*{!}{18mm}{\hfill\includegraphics{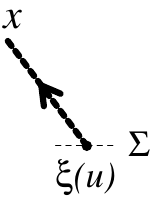}}}
\begin{equation}
\big[\xi\cdot{\mathbbm T}^{\vartheta}_\u\big]\raise -8mm\box1
=\;
\raise -8mm\box2\; .
\end{equation}
In words, $\xi\cdot{\mathbbm T}^{\vartheta}_\u$ replaces at the
point $u\in\Sigma$ the initial value of $\vartheta$ by $\xi(u)$.
Thanks to this operator, the perturbation to $\vartheta^b(x)$ that
results from a perturbation $\xi^b(u)$ of its initial condition on
the Cauchy surface $\Sigma$ can be written as
\begin{equation}
\xi^b(x)
=
\int\limits_\Sigma d^3\vec{\u}\;
[\xi\cdot{\mathbbm T}^{\vartheta}_\u]\;\vartheta^b(x)\; .
\end{equation}

When applying this formalism to the spectrum of the $\theta$ field, we
must remember that this spectrum does not depend on a single
$\vartheta$ function, but on a an infinite collection of such
functions, the $\vartheta_{\pm \q a}$, for all $\q$ and $a$. Therefore,
we need an operator such as the one defined in eq.~(\ref{eq:T-theta})
for each of them. To avoid the proliferation of notations, we will
continue to use the compact notation $\xi\cdot{\mathbbm
  T}^{\vartheta}_\u$ to denote the sum of these operators for all the
relevant fields,
\begin{equation}
\xi\cdot{\mathbbm T}^{\vartheta}_\u
\equiv
\sum_a \sum_{\epsilon=\pm}
\int {d^3\q}\;
\Bigg[
\xi^b_{\epsilon\q a}(u)\frac{\delta}{\delta\vartheta^b_{\epsilon\q a}(u)}
+(n\cdot\partial\xi^b_{\epsilon\q a}(u))\frac{\delta}{\delta(n\cdot\partial\vartheta^b_{\epsilon\q a}(u))}
\Bigg]
\; .
\end{equation}

In order to generate the missing terms in eq.~(\ref{eq:NLO-all}) (at
the exception of those corresponding to scalar vacuum fluctuations),
let us introduce the following two objects
\setbox1\hbox to 5.4cm{\hfil\resizebox*{!}{2cm}{\includegraphics{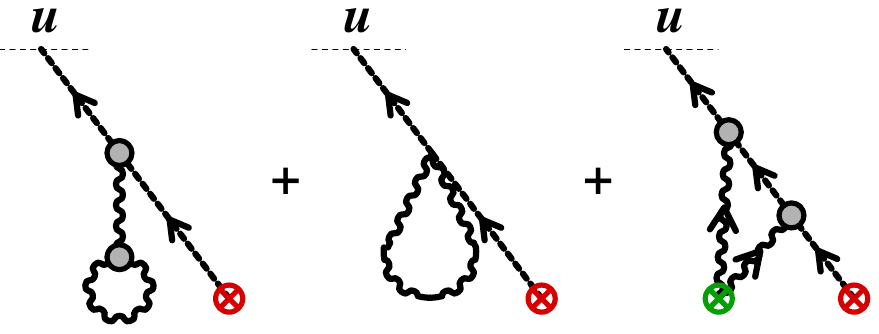}}\hfil}
\setbox2\hbox to 1.8cm{\hfil\resizebox*{!}{2cm}{\includegraphics{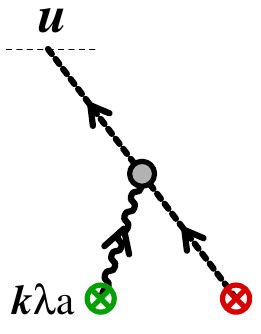}}\hfil}
\begin{eqnarray}
\beta(u)&\equiv&\raise -8mm\box1\; ,\nonumber\\
\zeta_{\k\lambda a}(u)&\equiv&\raise -9mm\box2\; .
\end{eqnarray}
It is now possible to generate the missing terms as follows~:
\begin{itemize}
\item[{\bf i.}] The operator $\beta\cdot{\mathbbm T}^{\vartheta}$ will
  generate the terms of eq.~(\ref{eq:NLO-all}) where the two vertices
  $u$ and $v$ are attached to the same scalar line and are both below
  the surface $\Sigma$.
\item[{\bf ii.}] The operator $\frac{1}{2}[\zeta_{-\k\lambda
    a}\cdot{\mathbbm T}^{\vartheta}][\zeta_{+\k\lambda
    a}\cdot{\mathbbm T}^{\vartheta}]$ generates the part of the term
  [{\bf i}] of eq.~(\ref{eq:NLO-all}), where the vertices $u$ and $v$
  are both below $\Sigma$.
\item[{\bf iii.}] In order to produce the part of the terms [{\bf
    e,g,i}] where one of the vertices $u$ and $v$ is above $\Sigma$
  while the other is below, we must act with the operator
  \begin{equation}
    \frac{1}{2}[a_{-\k\lambda
    a}\cdot{\mathbbm T}][\zeta_{+\k\lambda
    a}\cdot{\mathbbm T}^{\vartheta}]
  +\frac{1}{2}[\zeta_{-\k\lambda
    a}\cdot{\mathbbm T}^{\vartheta}][a_{+\k\lambda
    a}\cdot{\mathbbm T}]\; ,
  \end{equation}
  that mixes the gluonic and the scalar shifts.
\end{itemize}

\subsubsection{Final form of the NLO spectrum}
Let us now summarize this section by collecting all the terms
together. At the exception of the contribution of the scalar vacuum
fluctuations, the NLO correction to the $\theta$ inclusive spectrum
can be written as follows,
\begin{eqnarray}
&&
\left.\frac{dN_\theta}{d^3\p}\right|_{_{\rm NLO}}
=
\Bigg[\int\limits_\Sigma d^3\vec\u\,
\big[\alpha\cdot{\mathbbm T}_\u+\beta\cdot{\mathbbm T}_\u^\vartheta\big]
+
\nonumber\\
&&
+\frac{1}{2}\sum_{\lambda,a}\int\frac{d^3\k}{(2\pi)^3 2k}
\int\limits_\Sigma d^3\vec\u d^3\vec\v\,
\big[a_{-\k\lambda a}\cdot{\mathbbm T}_\u+\zeta_{-\k\lambda a}\cdot{\mathbbm T}_\u^\vartheta\big]
\big[a_{+\k\lambda a}\cdot{\mathbbm T}_\v+\zeta_{+\k\lambda a}\cdot{\mathbbm T}_\u^\vartheta\big]
\Bigg]\,
\left.\frac{dN_\theta}{d^3\p}\right|_{_{\rm LO}}\; .
\nonumber\\
&&
\label{eq:theta-NLO}
\end{eqnarray}
This formula is the main result of this paper.  It generalizes the
eqs.~(\ref{eq:g-NLO}) and (\ref{eq:phi-NLO}) to the case of a theory
involving new colored fields in addition to the gluons.  The
general structure of the operator that relates the NLO correction to
the LO is preserved by this new formula, and the generalization simply
amounts to introducing new terms that involve derivatives with
respect to the initial value of the scalar fields.

Although we have derived it here for scalar fields, a structurally
identical formula can be derived for quark production. The NLO
topologies in eq.~(\ref{eq:NLO-all}) are the same for the quark
spectrum, at the exception of $[{\bf b,d}]$ that do not exist for
quarks due to the absence of $q\bar{q}gg$ coupling. We have sketched
the derivation of the quark analogue of eq.~(\ref{eq:theta-NLO}) in
the appendix \ref{app:quarks}.

\subsection{Discussion on factorization}
In the case of the gluon spectra, the identity (\ref{eq:g-NLO}) was
essential in order to extract the logarithms of the CGC cutoff
$\Lambda$, and to prove that these logarithms can be factorized into
distributions of color sources that evolve according to the JIMWLK
equation. The reason for this is that, if one chooses the Cauchy
surface $\Sigma$ to be just above the past light-cone, then all these
logarithms are contained in the operator that relates LO and NLO, as
stated by the equation (\ref{eq:NLO-logs}). From this formula,
factorization follows easily thanks to the Hermiticity of the JIMWLK
Hamiltonian (see \cite{GelisLV3} for details).

Obviously, the operator in the left hand side of
eq.~(\ref{eq:NLO-logs}) is contained in the operator that appears in
the right hand side of eq.~(\ref{eq:theta-NLO}), and as a consequence
we already know that the NLO $\theta$ spectrum contains logarithms
proportional to the JIMWLK Hamiltonian. However, it is crucial to note
that we do not get the complete Hamiltonian here: we get the
JIMWLK Hamiltonian restricted to act only on the color sources
contained in the gluon fields on $\Sigma$. This is because it is
derived from the operators ${\mathbbm T}$, that act only on the
initial gauge fields, but not on the scalar fields. In other words, it
would be more correct to write\begin{eqnarray}
&&
\int\limits_\Sigma \!\!d^3\vec\u\,
\big[\alpha\cdot{\mathbbm T}_\u\big]
+
\frac{1}{2}\sum_{\lambda,a}\!\!\int\frac{d^3\k}{(2\pi)^3 2k}\!
\int\limits_\Sigma \!\!d^3\vec\u d^3\vec\v\,
\big[a_{-\k\lambda a}\cdot{\mathbbm T}_\u\big]
\big[a_{+\k\lambda a}\cdot{\mathbbm T}_\v\big]
=\nonumber\\
&&
\qquad\qquad\qquad\qquad\qquad
=
\log(\Lambda^+)\,{\cal H}_1^{_A}
+
\log(\Lambda^-)\,{\cal H}_2^{_A}
+\mbox{terms w/o logs}\; ,
\label{eq:NLO-logs-1}
\end{eqnarray}
where ${\cal H}_{1,2}^{_A}$ is this restricted JIMWLK Hamiltonian that
acts only on gluons. 

For factorization to be valid for scalar production, we would also
need the missing parts of the Hamiltonian, that act on the initial
scalar fields. This is presumably the role of the additional terms in
the operator that appears in eq.~(\ref{eq:theta-NLO}). Indeed, these
terms involve the new shift operator ${\mathbbm T}^\vartheta$, that
precisely acts on the initial value of the scalar fields. We are thus
led to conjecture an extension of eq.~(\ref{eq:NLO-logs}), that reads
\begin{eqnarray}
&&\int\limits_\Sigma d^3\vec\u\,
\big[\alpha\cdot{\mathbbm T}_\u+\beta\cdot{\mathbbm T}_\u^\vartheta\big]
+\nonumber\\
&&
\qquad
+\frac{1}{2}\sum_{\lambda,a}\int\frac{d^3\k}{(2\pi)^3 2k}
\int\limits_\Sigma d^3\vec\u d^3\vec\v\,
\big[a_{-\k\lambda a}\cdot{\mathbbm T}_\u+\zeta_{-\k\lambda a}\cdot{\mathbbm T}_\u^\vartheta\big]
\big[a_{+\k\lambda a}\cdot{\mathbbm T}_\v+\zeta_{+\k\lambda a}\cdot{\mathbbm T}_\u^\vartheta\big]
=
\nonumber\\
&&
\qquad
=\log(\Lambda^+)\,{\cal H}_1
+
\log(\Lambda^-)\,{\cal H}_2
+\mbox{terms w/o logs}\; ,
\label{eq:NLO-logs-2}
\end{eqnarray}
where now ${\cal H}_{1,2}$ is the full JIMWLK Hamiltonian that acts
both on the color sources contained in gauge fields and in scalar
fields. This formula, if true, would immediately imply the
factorizability of the leading logarithms of the $\theta$ spectrum
into the universal distributions $W$. However, proving
eq.~(\ref{eq:NLO-logs-2}) requires extracting the logarithms coming
from the new terms of eq.~(\ref{eq:theta-NLO}) and showing that their
coefficients recombine nicely to form the kernel of the JIMWLK
Hamiltonian. We stress that this has not been done at the moment.

\section{Summary and outlook}
\label{sec:summary}
In this paper, we have derived generalizations of an identity that
plays a crucial role in the proof of factorization of high energy
logarithms in heavy ion collisions. This functional relationship, that
relates the leading order and next-to-leading order contributions to
inclusive observables, is a central step in factorization because it
highlights the causal aspects of a collision process, by separating
what happens before the collision from what happens after.

The generalization we have obtained goes in the direction of extending
the field content of the theory under consideration. To the gluons
already present in all the known examples of this factorization, we
have added scalar fields that couple in some way to the gluons. The
simplest extension is that of color neutral scalar particles. Because
they are not colored, the formula that was already known for gluons is
valid without change in this situation.

Then, we have considered a much more complicated example, involving
colored scalar particles, that interact non trivially with the gauge
fields. In this case, the formula that was valid for the pure glue
case must be slightly modified. We have shown that this can be done by
a very natural extension of the formalism: we also need operators that
differentiate with respect to the initial value of the scalar fields.
Thanks to this generalization, it is also possible to obtain a
functional relationship between the LO and NLO contributions.

An obvious extension of this work is to use eq.~(\ref{eq:theta-NLO})
in order to extract the logarithms of the CGC cutoffs, and to prove
that they can be absorbed into the JIMWLK evolved $W$ distributions
that already appear in purely gluonic observables, providing further
evidence of their universality. Naturally, this would be most
interesting in the case of the quark inclusive spectrum, for which we have
outlined the relevant NLO formula in the appendix \ref{app:quarks}.

A more remote application of the same ideas and tools is to study the
logarithms that arise in gluon production beyond leading logarithmic
accuracy, in order to obtain the first $\alpha_s$ correction to the
JIMWLK Hamiltonian\footnote{The $\alpha_s$ correction to the (much
  simpler) Balitsky-Kovchegov equation --a mean field approximation of
  the JIMWLK evolution-- has been derived in
  \cite{Balit3,BalitC1,KovchW1,GardiKRW1}.}. The first step in this
program is to derive the inclusive gluon spectrum at NNLO, i.e. at two
loops. Some of the topologies one would need to evaluate in this
calculation are very similar to the graphs we have studied in the
section \ref{sec:scal2}. If one wants to extend the factorization
proof of \cite{GelisLV3} beyond leading logarithmic accuracy, an
important intermediate step would be to find a functional relationship
between the gluon spectra at NNLO and LO, similar to
eq.~(\ref{eq:g-NLO}).

There is another, much more intriguing and difficult, direction in
which one could try to extend these factorization results -- that of
exclusive observables, e.g. the cross section for events with a
rapidity gap. In the case of exclusive observables, the strategy for
establishing factorization that we have employed in \cite{GelisLV3}
fails at the first stage: so far we have not been able to obtain
formulae such as eqs.~(\ref{eq:g-NLO}), (\ref{eq:phi-NLO}) and
(\ref{eq:theta-NLO}) for any exclusive observable. The main obstacle
in obtaining similar formulae in this case seems to be that these
exclusive observables are expressible in terms of classical fields
with {\sl non-retarded} boundary conditions \cite{GelisV3} (for
instance boundary conditions that constrain the fields both at
$x^0=-\infty$ and at $x^0=+\infty$).

\section*{Acknowledgements}
This work is supported by the Agence Nationale de la Recherche project
\#~11-BS04-015-01. FG would like to thank G. Beuf, T. Lappi, T. Liou,
A.H. Mueller and R. Venugopalan for useful discussions on closely
related questions.

\appendix

\section{Inclusive quark spectrum}
\label{app:quarks}
The calculation of the inclusive spectrum at LO and NLO for charged
scalar fields, that we have presented in the section \ref{sec:scal2},
can easily be generalized to the case of quark production. Now the
Lagrangian density that couples the quarks and antiquarks to the
gluons is
\begin{equation}
{\cal L}\equiv i\overline{\psi}\,\slD\, \psi\; ,
\end{equation}
where $\psi$ is the spinor for one quark flavor (we disregard the
quark mass, that does not play any role in this discussion).  This
Lagrangian is quadratic in the fermion fields, which means that the
quarks do not interact directly among themselves.  In terms of its
interactions, this theory is in fact a bit simpler than the theory of
charged scalars, since it has only a $q\overline{q} g$ vertex and no
derivative couplings.

The inclusive quark spectrum is given to all orders by a formula which
is very similar to eq.~(\ref{eq:theta-allO1}),
\begin{equation}
\frac{dN_{\rm q}}{d^3\p}
=
\frac{1}{(2\pi)^3 2p}\sum_{s,i}\int d^4x d^4y\; e^{ip\cdot(x-y)}\;
(u_{s}(\p)\slpartial_x)(u^\dagger_{s}(\p)\slpartial_y)\;S_{+-}^{ii}(x,y)\; ,
\label{eq:quark-allO1}
\end{equation}
where $u_s(\p)$ is a free spinor representing a quark of momentum $\p$
and spin $s$, and $S_{+-}^{ij}(x,y)$ is the $+-$ component of the
quark two-point function in the Schwinger-Keldysh formalism (we use
latin letters $i,j,k$ for colors indices in the fundamental
representation). Note that the inclusive spectrum is summed over all
spin states $s$ and colors $i$ of the produced quark.

At leading order, we can represent this two-point function in terms of a
complete basis of solutions of the Dirac equation in the background
field ${\cal A}^\mu$,
\begin{eqnarray}
  {\cal S}_{+-}^{ij}(x,y)
  &=&
  \sum_{s,k}\int\frac{d^3\q}{(2\pi)^3 2q}\;
  \psi^{\dagger i}_{\q s k}(x)\psi^j_{\q s k}(y)
  \nonumber\\
  i\slcalD_{ij}\,\psi^j_{\q s k}(x) &=& 0\; ,\quad 
\lim_{x^0\to -\infty}\psi^i_{\q s k}(x) = \delta_{ik}\, v_s(\q)\, e^{i q \cdot x}\; .
\label{eq:S+--LO}
\end{eqnarray}
Here also, the lower color index is the color of the quark at
$x^0=-\infty$, while $i$ is its color at the current time $x^0$.
Eqs.~(\ref{eq:quark-allO1}) and the representation (\ref{eq:S+--LO})
were the basis of the numerical evaluation of the quark spectrum in
the CGC framework done in \cite{GelisKL1,GelisKL2}.

\begin{figure}[htbp]
\begin{center}
\resizebox*{!}{1.7cm}{\includegraphics{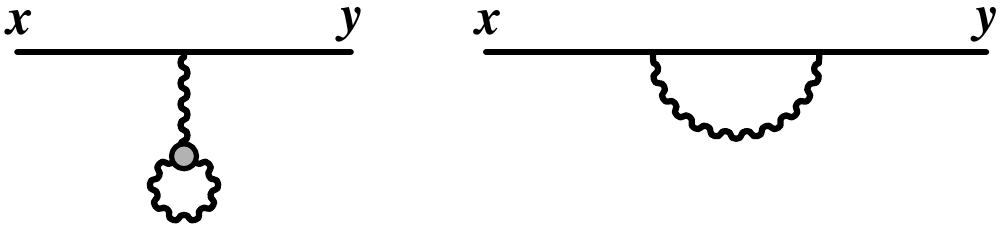}}
\end{center}
\caption{\label{fig:q-NLO}Diagrammatic representation of the
  contributions to the inclusive quark spectrum at Next-to-Leading
  Order.  All the lines represent propagators dressed by insertions of
  the classical color field ${\cal A}^\mu$. The blobs indicate that
  the corresponding vertices are dressed by the background field.}
\end{figure}

At NLO, we need to evaluate the 1-loop corrections to the two-point
function $S_{+-}$. The relevant topologies are those listed in the
figure \ref{fig:q-NLO}.
One of the topologies present in the scalar case does not exist for quarks,
since there is no $q\overline{q}gg$ elementary vertex. For the same
reason, the $q\overline{q}g$ vertex cannot be dressed by the
background field. After some transformations identical to those
performed in the section \ref{sec:alt}, we can represent the NLO
contributions to the quark spectrum in the following form, which is
more appropriate for our purposes:
\setbox1\hbox to 13cm{\hfil\resizebox*{!}{4.0cm}{\includegraphics{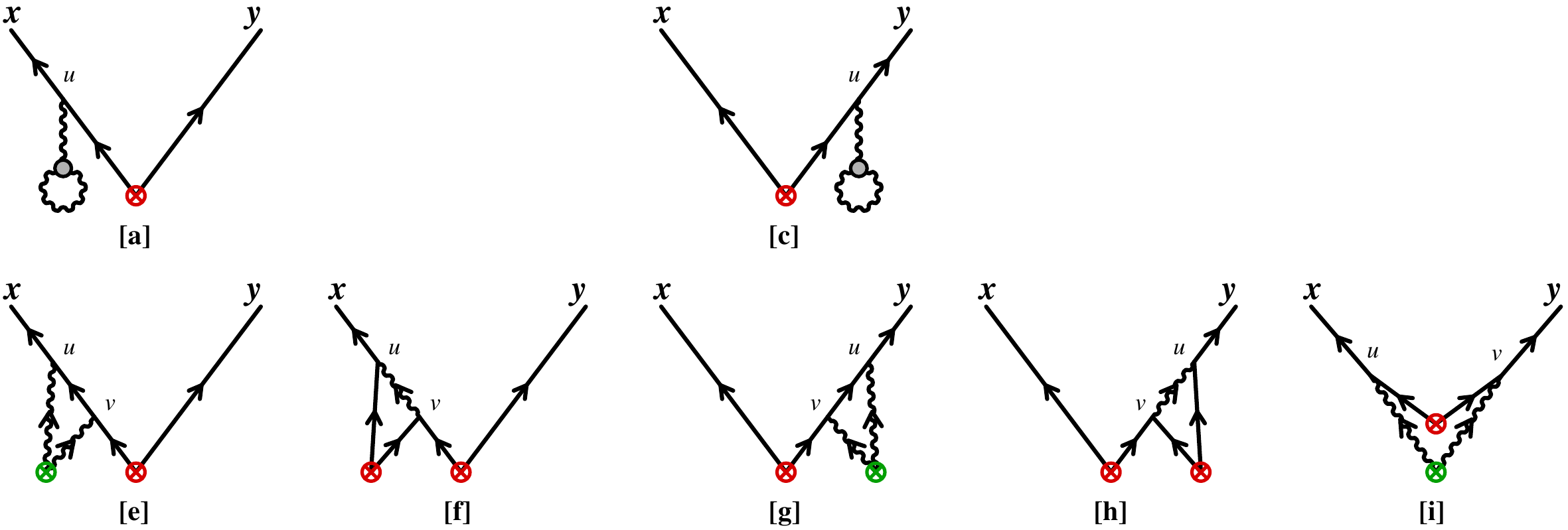}}}
\begin{equation}
\raise -10mm\box1\quad .
\label{eq:q-NLO-all}
\end{equation}
Compared to eq.~(\ref{eq:NLO-all}), two graphs have disappeared, and
all the $q\overline{q}g$ vertices are bare vertices.  Like in the
scalar case, all the NLO terms that involve gluonic vacuum
fluctuations (i.e. all the above terms except the graphs [{\bf f,h}])
can be related to the LO spectrum as follows,
\begin{eqnarray}
&&
\left.\frac{dN_{\rm q}}{d^3\p}\right|_{_{\rm NLO}}
=
\Bigg[\int\limits_\Sigma d^3\vec\u\,
\big[\alpha\cdot{\mathbbm T}_\u+\beta\cdot{\mathbbm T}_\u^{\rm Q}\big]
+
\nonumber\\
&&
+\frac{1}{2}\sum_{\lambda,a}\int\frac{d^3\k}{(2\pi)^3 2k}
\int\limits_\Sigma d^3\vec\u d^3\vec\v\,
\big[a_{-\k\lambda a}\cdot{\mathbbm T}_\u+\zeta_{-\k\lambda a}\cdot{\mathbbm T}_\u^{\rm Q}\big]
\big[a_{+\k\lambda a}\cdot{\mathbbm T}_\v+\zeta_{+\k\lambda a}\cdot{\mathbbm T}_\u^{\rm Q}\big]
\Bigg]\,
\left.\frac{dN_{\rm q}}{d^3\p}\right|_{_{\rm LO}}\; .
\nonumber\\
&&
\label{eq:quark-NLO}
\end{eqnarray}
All the differences between adjoint scalars and quarks are now hidden
in the definitions of the objects $\beta$ and $\zeta_{\pm\k\lambda
  a}$, that are now given by
\setbox1\hbox to 4.0cm{\hfil\resizebox*{!}{2cm}{\includegraphics{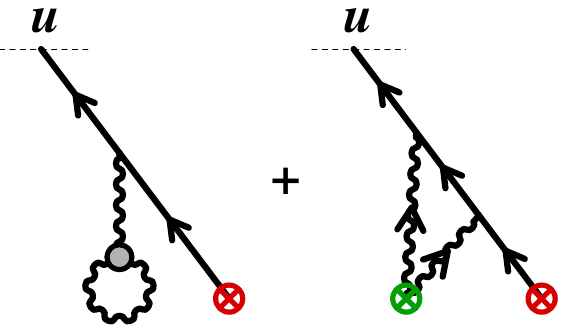}}\hfil}
\setbox2\hbox to 1.8cm{\hfil\resizebox*{!}{2cm}{\includegraphics{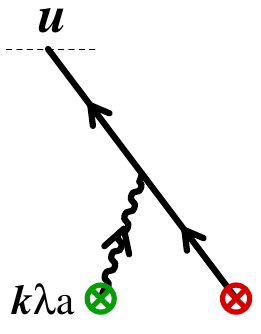}}\hfil}
\begin{equation}
\beta(u)\equiv\raise -8mm\box1\; ,\quad
\zeta_{\k\lambda a}(u)\equiv\raise -9mm\box2\; ,
\end{equation}
and in the definition of the operator ${\mathbbm T}_\u^{\rm Q}$ that
replaces ${\mathbbm T}_\u^\vartheta$. This new operator generates
shifts of the initial condition of a spinor on the Cauchy surface
$\Sigma$, and eq.~(\ref{eq:T-theta}) is replaced by
\begin{equation}
\xi\cdot{\mathbbm T}^{\rm Q}_\u
\equiv
(\sln\xi^i(u))\frac{\delta}{\delta(\sln\psi^i(u))}
\; .
\label{eq:T-quark}
\end{equation}



\end{document}